\documentclass[apj,superscriptaddress]{emulateapj}

\usepackage{amsmath}
\usepackage{natbib}
\usepackage{epsfig}
\usepackage{txfonts}

\newcommand{\id}{{\,\rm d}}

\newcommand{\beq}{\begin{equation}}   %

\newcommand{\eeq}{\end{equation}}   %

\newcommand{\beqa}{\begin{eqnarray}}   %

\newcommand{\eeqa}{\end{eqnarray}}   %

\newcommand{\beal}{\begin{align}}

\newcommand{\enal}{\end{align}}

\newcommand{\bspl}{\begin{split}}

\newcommand{\espl}{\end{split}}

\newcommand{\bsub}{\begin{subequations}}

\newcommand{\esub}{\end{subequations}}

\newcommand{\bmulti}{\begin{multline}}   %

\newcommand{\beqm}{\begin{mathletters}}   %

\newcommand{\eeqm}{\end{mathletters}}   %

\newcommand{\Ne}{N_{\rm e}}

\newcommand{\sigT}{\sigma_{\rm T}}

\newcommand{\pot}[2]{#1 \times 10^{#2}}

\newcommand{\HeIlevel}[4]{{#1^{#2} {\rm #3}_{#4}}}   

\newcommand{\zmu}{{z_{\mu}}}
\newcommand{\zmuy}{{z_{\mu,y}}}

\newcommand{\nS}{n_{\rm S}}

\newcommand{\nrun}{n_{\rm run}}

\newcommand{\taudot}{\dot{\tau}}
\newcommand{\taudotc}{\tau'}
\newcommand{\kD}{k_{\rm D}}
\newcommand{\rs}{r_{\rm s}}
\newcommand{\cs}{c_{\rm s}}

\usepackage{color}

\usepackage{hyperref}
\usepackage{grffile}
\usepackage{graphics}
\usepackage{booktabs}

\slugcomment{Draft version \today}
\shorttitle{Power spectrum constraints}
\shortauthors{Chluba, Erickcek \& Ben-Dayan}

\begin{document}

\title{Probing the inflaton: Small-scale power spectrum constraints \\ from measurements of the CMB energy spectrum}

\author{Jens Chluba\altaffilmark{1},
Adrienne L. Erickcek\altaffilmark{1,2}
and 
Ido Ben-Dayan\altaffilmark{1,2}
}
\altaffiltext{1}{Canadian Institute for Theoretical Astrophysics, 60 St. George Street,
Toronto, Ontario M5S 3H8, Canada}
\altaffiltext{2}{Perimeter Institute for Theoretical Physics, 31 Caroline St. N, 
Waterloo, Ontario N2L 2Y5, Canada}

\email[Please direct questions to ]{jchluba@cita.utoronto.ca}

\begin{abstract}
In the early Universe, energy stored in small-scale density perturbations is quickly dissipated by Silk-damping, a process that inevitably generates $\mu$- and $y$-type spectral distortions of the cosmic microwave background (CMB).
These spectral distortions depend on the shape and amplitude of the primordial power spectrum at wavenumbers $k\lesssim 10^4\,{\rm Mpc}^{-1}$.
Here we study constraints on the primordial power spectrum derived from COBE/FIRAS and forecasted for PIXIE.
We show that measurements of $\mu$ and $y$ impose strong bounds on the integrated small-scale power, and we demonstrate how to compute these constraints using $k$-space window functions that account for the effects of thermalization and dissipation physics.
We show that COBE/FIRAS places a robust upper limit on the amplitude of the small-scale power spectrum. This limit is about three orders of magnitude stronger than the one derived from primordial black holes in the same scale range. Furthermore, this limit could be improved by another three orders of magnitude with PIXIE, potentially opening up a new window to early Universe physics.
To illustrate the power of these constraints, we consider several generic models for the small-scale power spectrum predicted by different inflation scenarios, including 
running-mass inflation models and inflation scenarios with episodes of particle production. 
PIXIE could place very tight constraints on these scenarios, potentially even ruling out running-mass inflation models if no distortion is detected.
We also show that inflation models with sub-Planckian field excursion that generate detectable tensor perturbations should simultaneously produce a large CMB spectral distortion, a link that could potentially be established by PIXIE.
\end{abstract}




\section{Introduction}
\label{sec:Intro}
{Cosmological inflation \citep{AS82,Guth80,Linde82} provides a commonly accepted} explanation for both the Universe's homogeneity and the origin of the initial curvature perturbations that seeded the growth of structure.  Inflation cannot be considered a complete theory, however, until we understand the inflaton: the field that drove an epoch of accelerated expansion in the early Universe.  Fortunately, the statistical properties of the initial density perturbations offer a wealth of information about inflationary physics.   In single field inflation, we can in principle reconstruct the inflaton potential if the primordial power spectrum is known at all scales (e.g. \citealt{LLK97}).
However, the limited range of scales probed by the CMB and large scale structure (LSS) does not provide sufficient information to discriminate between many inflation models. Finding additional ways to measure the primordial power spectrum outside this range of scales will greatly enhance our ability to constrain the inflaton's potential and its trajectory during inflation. In this work, we investigate how spectral distortions of the CMB caused by the dissipation of energy stored in small-scale density perturbations can provide a new probe of inflation by extending our knowledge of the primordial power spectrum from $k\simeq 1\,{\rm Mpc}^{-1}$ to about $10^4\,{\rm Mpc}^{-1}$.

The simplest models of inflation predict a power spectrum parameterized by a nearly constant, slightly red spectral index. 
More complicated inflationary models can leave distinctive imprints in the primordial power spectrum.  Multi-field inflation can produce primordial power spectra with steps \citep{1987PhRvD..35..419S, Polarski:1992dq, 1997NuPhB.503..405A} or oscillations \citep{2011JCAP...01..030A, Takeshi2011, 2012arXiv1201.4848C}.  These features in the primordial power spectrum may also be generated during single-field inflation by discontinuities, kinks, and bumps in the inflaton potential \citep{1989PhRvD..40.1753S, 1992JETPL..55..489S, 1994PhRvD..50.7173I, 1998GrCo....4S..88S, Hunt:2007dn, 2008PhRvD..77b3514J}.  The primordial power spectrum may also contain information about how the inflaton interacts with other fields; for instance, particle production during inflation leaves a bump in the primordial power spectrum \citep{2000PhRvD..62d3508C, Neil2009I, Barnaby2010}.  
Finally, several inflationary models predict enhancement of the small-scale perturbations that are generated during the later stages of inflation \citep{1996NuPhB.472..377R, Stewart1997b, 1998PhRvD..58f3508C, 1999PhRvD..59f3515C, 1999PhRvD..60b3509C, 2000PhRvD..61h3518M, 2001PhRvD..63l3501M, 2010JCAP...09..007B, 2011JCAP...03..028G, 2011JCAP...07..035L,2011JCAP...11..028B}.

The CMB temperature fluctuations provide a precise measurement of the primordial power spectrum on large scales, corresponding to wavenumbers $10^{-3} \, {\rm Mpc}^{-1}\lesssim k \lesssim 0.1\,{\rm Mpc}^{-1}$ \citep{CBI03, ACBAR09, Quad09, Larson2011, ACT11}.  
Luminous red galaxies and galaxy clusters probe the matter power spectrum on similar scales \citep[$0.02\, {\rm Mpc}^{-1}\lesssim k \lesssim 0.7\, {\rm Mpc}^{-1}$;][]{SDSSDR7, VKB09, TSW11, STA11}, while the Lyman-$\alpha$ forest reaches slightly smaller scales \citep[$0.3\, {\rm Mpc}^{-1} \lesssim k \lesssim 3 \, {\rm Mpc}^{-1}$;][]{MSB06}.  All these observations indicate that the primordial power spectrum is nearly scale-invariant with an amplitude {close} to $\pot{2}{-9}$ \citep{TZ02, NC09, Komatsu2010, Dunkley2010, Keisler2011, ACT11, BPVV11}. There is no evidence of features in the primordial power spectrum on these scales \citep{Kinney:2008wy, Mortonson:2009qv, Neil2009, Hamann:2009bz, Peiris:2009wp, Dvorkin2010, Bennett:2010jb, Benetti:2011rp, Dvorkin2011}.

Our knowledge of the primordial power spectrum on smaller scales is far more limited; we only have upper bounds on its amplitude for $k\gtrsim 3\,{\rm Mpc}^{-1}$.  One of these upper bounds is derived from the limits on spectral distortions in the CMB.
It was long understood that the Silk-damping \citep{Silk1968} of primordial small-scale perturbations causes energy release in the early Universe \citep{Sunyaev1970diss, Daly1991, Barrow1991, Hu1994}. 
This gives rise to small spectral distortions of the CMB spectrum that directly depend on the shape and amplitude of the primordial power spectrum.
Modes with wavenumbers $50\,{\rm Mpc^{-1}}\lesssim k \lesssim 10^4\,{\rm Mpc^{-1}}$ dissipate their energy during the $\mu$-era (redshift $\pot{5}{4}\lesssim z \lesssim \pot{2}{6}$), producing a non-vanishing constant residual chemical potential at high frequencies \citep{Sunyaev1970mu, Zeldovich1972, Illarionov1974, Burigana1991, Hu1993}, while modes with $k \lesssim 50\,{\rm Mpc^{-1}}$ result in a $y$-distortion. The latter is also well known in connection with the SZ-effect of clusters of galaxies \citep{Zeldovich1969}.
By accurately measuring the CMB spectrum one can therefore place robust upper limits on the possible power at small scales since the physics going into the production of these distortions is well understood.

Very precise measurements of the CMB spectrum were obtained with COBE/FIRAS \citep{Mather1994, Fixsen1996}, limiting possible deviations from a blackbody to $\mu\lesssim\pot{9}{-5}$ and $y \lesssim\pot{1.5}{-5}$ at 95\% confidence \citep{Fixsen1996}.
At lower frequencies, a similar limit on $\mu$ was recently obtained by ARCADE \citep{arcade2}, and $\mu\lesssim\pot{6}{-5}$ at $\nu \simeq 1\,{\rm GHz}$ is derived from TRIS \citep{tris1, tris2}.
For power spectra with constant spectral index, $\nS$, and normalization fixed at CMB scales, the measurements of COBE/FIRAS imply $\nS \lesssim 1.6$ \citep{Hu1994}, 
but this limit is model-dependent. For instance, a small negative running, $\nrun$, of the spectral index weakens this bound significantly \citep{Khatri2011, Chluba2012}.

Here we generalize the COBE/FIRAS limits on spectral distortions by directly converting them into a bound on the total perturbation power at small scales. Depending on the particular inflationary model, this translates into constraints on different model parameters; the conversion can be obtained on a case-by-case basis.
Also, the recently proposed CMB experiment PIXIE \citep{Kogut2011PIXIE} might be able to detect distortions that are $\sim 10^3$ smaller than the upper limits given by COBE/FIRAS.
At this level of sensitivity, PIXIE is already close to what is required to detect the distortions arising  from
the dissipation of acoustic modes for a power spectrum with $\nS=0.96$ and no running  all the way from CMB-anisotropy scales to $k\simeq 10^4$ \citep{Chluba2011therm, Khatri2011, Chluba2012}.
Such an improvement could rule out inflationary models with additional power at small scales, as we discuss here in more detail. 
%
%
Conversely, any detection of spectral distortions implies that either the power spectrum is enhanced on small scales, contrary to the predictions of the simplest inflation models, or an alternative mechanism generated CMB spectral distortions in the early Universe (e.g. particle decays).

The only other upper bounds on the amplitude of the small-scale primordial power spectrum are derived from the absence of primordial black holes (PBHs) and ultracompact minihalos (UCMHs), which are dense dark matter halos {that form} at high redshift ($z \simeq 1000$).  Both PBHs and UCMHs form in regions with large primordial overdensities; an initial overdensity of $\delta \rho/\rho \gtrsim 0.3$ is required to form a PBH \citep{Carr75, NJ99}, while UCMHs form in regions where $\delta \rho/\rho \gtrsim 10^{-3}$ when they enter the Hubble horizon \citep{RG09, BSA11}.  
There are numerous constraints on the number density of PBHs; \citet{JGM09} showed that these constraints imply that the amplitude of the primordial curvature power spectrum is less than 0.01-0.06 over an extremely wide range of scales ($0.01\,{\rm Mpc^{-1}}\lesssim k \lesssim 10^{23}\,{\rm Mpc^{-1}}$).  Even though PBHs provide only a weak upper bound on the small-scale amplitude of the primordial power spectrum, they have usefully constrained inflationary models \citep[e.g.,][]{1993PhRvD..48..543C,2000PhRvD..62d3516L,2008JCAP...04..038K,2008JCAP...07..024P,2010PhRvD..82d7303J, 2011arXiv1107.1681L, 2011arXiv1112.5601B}. 

Since UCMHs form in lower density regions than those that produce PBHs, limits on their abundance can provide tighter constraints on the primordial power spectrum \citep{JG10, BSA11}.  Unfortunately, all current limits on the number density of UCMHs rely on the assumption that they emit gamma rays from the annihilation of dark matter particles within their high-density centers \citep{SS09, BDEK10, LB10, YFH11, YHCZ11, YCL11, Zhang11, BSA11}.  If dark matter is a self-annihilating thermal relic, \citet{BSA11} recently showed that the Large Area Telescope on the Fermi Gamma-Ray Space Telescope \citep{Atwood2009} places the strongest constraint on the UCMH abundance; this limit implies that the amplitude of the primordial curvature power spectrum is less than $2\times10^{-7}$ to $2\times10^{-6}$ for modes with $10\,{\rm Mpc^{-1}}\lesssim k \lesssim 10^{7}\,{\rm Mpc^{-1}}$.  If dark matter does not self-annihilate, then UCMHs can only be detected gravitationally.  In this case, \citet{LEW12} recently showed that the Gaia satellite \citep{Lindegren2011} will be able to detect astrometric microlensing by UCMHs and that a null detection of UCMHs by Gaia would constrain the amplitude of the primordial power spectrum to be less than $10^{-5}$ for $k\simeq 3500\, {\rm Mpc}^{-1}$.

CMB spectral distortions probe the amplitude of the primordial power spectrum in a very different manner than PBHs and UCMHs.  
First of all, the physics underlying the computation of the CMB spectral distortions is very well understood, while, for example, the constraints derived from UCMHs depend on unknown properties of the dark matter particle: its mass, the abundance of its antiparticle, and its annihilation cross section.
Second, since they arise from overdense regions, PBHs and UCMHs probe the high-density tail of the probability distribution function for density perturbations.  The likelihood of forming a PBH or an UCMH is therefore highly sensitive to deviations from Gaussianity that enhance or suppress the abundance of high overdensities.  In contrast, the CMB spectral distortion is determined by the total energy stored in density perturbations and is therefore less sensitive to the precise form of the probability distribution function.  Furthermore, the expected number density of PBHs or UCMHs of a particular mass is determined by the mass variance in the sphere with radius $R$ that formed the PBH or UCMH, which is computed by convolving the primordial power spectrum with a filter function that is narrowly peaked at $kR\simeq1$.  Therefore, the abundance of UCMHs or PBHs of a particular mass probes the primordial power spectrum over a small range of scales.  Meanwhile, the CMB spectral distortion produced by the dissipation of acoustic modes depends on the amplitude of the primordial power spectrum over a much wider range of scales (e.g., $50\,{\rm Mpc^{-1}}\lesssim k \lesssim 10^4\,{\rm Mpc^{-1}}$ for $\mu$-distortions).  Consequently, limits on the power spectrum derived from CMB spectral distortions are more sensitive to the overall shape of the primordial power spectrum than those derived from the absence of UCMHs and PBHs.

Possible constraints on $\nS$ and $\nrun$  {for adiabatic perturbations} from future measurements of $\mu$ and $y$ with PIXIE are discussed in detail by \citet{Chluba2012}.
For a power spectrum with constant spectral index, PIXIE could independently rule out $\nS>1.05$ at $\sim 5\sigma$-level {and a pure Harrison-Zeldovich power spectrum at $\sim 2.5\sigma$-level}. This limit is driven mainly by the amount of small-scale power at wavenumbers $50\,{\rm Mpc^{-1}}\lesssim k \lesssim 10^4\,{\rm Mpc^{-1}}$, but a simple extrapolation from CMB-anisotropy scales down to these small scales is not necessarily correct.

Here we consider more generic models for the primordial power spectrum that match the measurements from the CMB and LSS on large scales but have enhanced power on smaller scales.
We begin in Section~\ref{sec:physics} by reviewing how the dissipation of small-scale inhomogeneities generates CMB spectral distortions, and we provide $k$-space window functions that facilitate the computation of the spectral distortion generated by a given power spectrum.  In Section~\ref{sec:constraints}, we compute the spectral distortions produced by three generic types of power spectra: power spectra with steps, kinks, and bumps.  Most inflationary models that predict excess power on small scales generate power spectra with these types of features, and the constraints derived in Section~\ref{sec:constraints} should be readily applicable to these models.
We also specifically discuss particle production during inflation (\S~\ref{sec:Neil}), running-mass inflation (\S~\ref{sec:running_mass}), and small-field inflation models ($\Delta\phi < M_{\rm Pl}$ during inflation; see \S~\ref{sec:Ido} for more details) that generate detectable gravitational waves.  We summarize our analysis in Section~\ref{sec:disc_conc}.

\section{Spectral distortions caused by the dissipation of acoustic modes}
\label{sec:physics}
A consistent microphysical treatment of the dissipation of acoustic modes in the early Universe was recently given by \citet{Chluba2012}.
There it was shown that temperature perturbations in the photon field set up by inflation lead to an average photon energy density $\left<\rho_\gamma\right>\approx \bar{\rho}^{\rm bb}_\gamma[1+6\left<\Theta^2 \right>]$ that in second order of the temperature fluctuations, $\Theta=\Delta T/\bar{T}$, is slightly larger than the energy density $\bar{\rho}^{\rm bb}_\gamma=a_{\rm R}\bar{T}^4$ of a blackbody at average photon temperature $\bar{T}$.
The temperature anisotropies at very small scales are subsequently completely erased by shear viscosity and thermal conduction \citep{WeinbergBook}, processes that isotropize the photon-baryon fluid. 
However, the energy stored in these perturbations of the medium is not lost but merely redistributed to larger scales, causing a small increase of the average photon temperature and resulting in an average spectral distortion by the mixing of blackbody spectra with different temperatures \citep{Zeldovich1972, Chluba2004}.
The effective energy release depends directly on the shape of the primordial power spectrum with 1/3 of the dissipated energy sourcing $y$-type spectral distortions that later thermalize, slowly approaching a $\mu$-type distortion.
The remaining 2/3 of energy just causes an adiabatic increase of the average photon temperature, without creating any distortion.

In \citet{Chluba2012} the photon Boltzmann equation for the average spectral distortion describing the effect of energy release from the dissipation of acoustic modes was derived and solved for primordial power spectra with constant spectral index and small running using the cosmological thermalization code {\sc CosmoTherm} \citep{Chluba2011therm}.
It was shown that once the source function, $\left<\mathcal{S}_{\rm ac}\right>$, for the primordial dissipation problem is known, a rather precise description of the resulting distortion can be obtained by computing the {\it weighted} energy release in the $\mu$- and $y$-era.
Given $\left<\mathcal{S}_{\rm ac}\right>$ the required effective energy release rate caused by the damping of acoustic modes is determined by
%
\beal
\label{eq:Q_ac}
\frac{1}{\rho_\gamma}\frac{\id Q_{\rm ac}}{\id z}
& = 
\frac{4\taudot \left<\mathcal{S}_{\rm ac}\right>}{H (1+z)},
\end{align}
where $\taudot =\sigT\Ne c \approx \pot{4.4}{-21}(1+z)^{3}\,{\rm sec^{-1}}$ denotes the rate of Thomson scattering and $H\approx \pot{2.1}{-20}\,(1+z)^2 {\rm sec^{-1}}$ is the Hubble expansion rate\footnote{
The approximations for $\taudot$ and $H$ are only valid at high redshifts during the radiation-dominated era.}.

Defining the {\it visibility function for spectral distortions}, $\mathcal{J}_{\rm bb}(z)=\exp\left(-[z/\zmu]^{5/2}\right)$, with $\zmu\approx \pot{1.98}{6}$, the weighted total energy release in the $\mu$- and $y$-era is
\bsub
\label{eq:def_mu_y_Dr_r}
\beal
\label{eq:def_mu_y_Dr_r_a}
\left.\frac{\Delta\rho_\gamma}{\rho_\gamma}\right|_{\mu}
&= 
\int_\zmuy^\infty  \frac{\mathcal{J}_{\rm bb}(z)}{\rho_\gamma}\,\frac{\id Q_{\rm ac}}{\id z} \id z
\\[1mm]
\label{eq:def_mu_y_Dr_r_b}
\left.\frac{\Delta\rho_\gamma}{\rho_\gamma}\right|_{y}
&= 
\int^\zmuy_0  \frac{1}{\rho_\gamma}\,\frac{\id Q_{\rm ac}}{\id z} \id z,
\end{align}
\esub
where $\zmuy\approx \pot{5}{4}$ \citep[cf.][]{Hu1993}.
At $z\gg\zmu$, the thermalization process is very efficient, so all the released energy just increases the specific entropy
of the Universe, and hence only raises the average temperature of the CMB without causing significant spectral distortions.
{However, below $\zmu$, the CMB spectrum becomes vulnerable, and energy release leads to spectral distortions.}
With the simple expressions from 
\citet{Sunyaev1970mu}, 
$\mu \approx 1.4 \,\Delta \rho_\gamma/\rho_\gamma|_\mu$ and $y \approx \frac{1}{4} \Delta \rho_\gamma/\rho_\gamma|_y$;
this can be used to estimate the expected residual distortion at high frequencies.

The energy release depends directly on the primordial power spectrum, $P_\zeta(k)$, of curvature perturbations.
Below we consider different cases for $P_\zeta(k)$, often parametrizing it as 
\beal
\label{eq:P_k}
P_\zeta(k)&=P^{\rm st}_\zeta(k)+\Delta P_\zeta(k)
\end{align}
where $\Delta P_\zeta(k)$ describes the deviation of the power spectrum from the commonly used form \citep{Kosowsky1995}, 
\beal
\label{eq:P_k_st}
P^{\rm st}_\zeta(k)&=2\pi^2 A_\zeta k^{-3} (k/k_0)^{\nS-1+\frac{1}{2} n_{\rm run} \ln(k/k_0)}
\end{align}
with $n_{\rm run}\equiv \id \nS / \id \ln k$. In the text we often refer to $P^{\rm st}_\zeta(k)$ as standard or background power spectrum.

From observations with WMAP at large scales we have
$A_\zeta=\pot{2.4}{-9}$ for pivot scale $k_0=0.002\,{\rm Mpc}^{-1}$ \citep{Komatsu2010, Dunkley2010, Keisler2011}.
%
%
Without running we have $\nS=0.963 \pm 0.014$ from WMAP7 only, while with running the currently favored values are $\nS=1.027\pm 0.051$ and $\nrun=-0.034 \pm 0.026$ \citep{Larson2011, Komatsu2010}.
More recent measurements of the damping tail of the CMB power spectrum by {ACT} \citep{Dunkley2010} and {SPT} \citep{Keisler2011} yield\footnote{For both experiments we quote the constraint derived in combination with $\rm WMAP7+BAO+H_0$.} $\nS=1.017\pm 0.036$ and $\nrun=-0.024 \pm 0.015$ and $\nS=0.9758\pm 0.0111$ and $\nrun=-0.020 \pm 0.012$, respectively.

As Eq.~\eqref{eq:def_mu_y_Dr_r} and our discussion below indicates, any constraint derived from $\mu$ or $y$-type distortions to leading order is determined by computing (i) the time average energy release over the redshift interval corresponding to the $\mu$ and $y$-era, and (ii) a weighted average of the total power stored over a particular range of scales. This means that the cosmological dissipation process provides a tight {\it integral constraint} on the power spectrum, which strongly limits possible inflaton trajectories in a very model-independent way.
Furthermore, this constraint is not limited to just inflation scenarios but in practice should be respected by any model invoked to create the primordial seeds of structures in our Universe.

\subsection{Computing the effective heating rate}
\label{sec:effective_heating_rate}
Here we are mainly interested in CMB spectral distortions caused by modes that dissipate most of their energy at redshifts well before the cosmological recombination epoch ($z\gtrsim 10^4$), when the Universe is still radiation-dominated and the baryon loading $R=3\rho_{\rm b}/4\rho_\gamma\approx 673 \, (1+z)^{-1}$ is negligible.
In this regime one can use the tight coupling approximation to compute the required source term for the photon Boltzmann equation \citep{Chluba2012}:
\beal
\left<\mathcal{S}_{\rm ac}\right>
& \approx 
\frac{\alpha_{\nu}}{\taudotc}\,\partial_\eta \kD^{-2}
\int \frac{\id^3 k}{(2\pi)^3}\,k^2 P_\zeta(k)\, 2\sin^2\left(k\rs\right)\, e^{-2k^2/\kD^2},
\nonumber
\end{align}
where $\alpha_{\nu}=[1+4R_\nu/15]^{-2}\approx 0.81$ with $R_\nu=\rho_\nu/(\rho_\gamma+\rho_\nu)\approx 0.41$ denoting the contributions of massless neutrinos to the energy density of relativistic species; 
$\eta=\int \frac{c\id t}{a}\approx \pot{4.7}{5}(1+z)^{-1}\,{\rm Mpc}$ denotes conformal time and $a=(1+z)^{-1}$ the scale factor normalized to unity today.
Furthermore, $\rs\approx \eta / \sqrt{3}\approx \pot{2.7}{5}(1+z)^{-1}\,{\rm Mpc}$ is the sound horizon; $\taudotc = a \sigT \Ne\approx \pot{4.5}{-7}(1+z)^{2}\,{\rm Mpc^{-1}}$ is the derivative of the Thomson optical depth with respect to $\eta$; $\kD\approx \pot{4.0}{-6}(1+z)^{3/2}\,{\rm Mpc^{-1}}$ determines the damping scale with \citep{Kaiser1983, Zaldarriaga1995}
\beal
\partial_\eta \kD^{-2}=
\frac{\cs^2}{2 \taudotc}\left[\frac{R^2}{1+R}+\frac{16}{15}\right]
\approx \frac{8}{45 \taudotc}\approx \pot{3.9}{5}(1+z)^{-2}\,{\rm Mpc}
\nonumber
\end{align}
and dimensionless sound speed $\cs=1/\sqrt{3(1+R)}$ of the tightly coupled photon-baryon fluid.
 {The above expression for $\left<\mathcal{S}_{\rm ac}\right>$ is based on the transfer functions for adiabatic perturbations, however, a similar formula can be obtained for isocurvature modes. Some discussion can be found in \citet{Hu1994isocurv} and \citet{Dent2012}.}

In the limit $R\ll 1$, the effective energy release rate for the photon field is therefore given by
\beal
\label{eq:Qdot_int}
\frac{1}{\rho_\gamma}\frac{\id Q_{\rm ac}}{\id z}
&\approx 
\frac{32\alpha_{\nu}c}{45 \taudotc H}
\int \frac{\id k}{2\pi^2}\,k^4 P_\zeta(k)\, 2\sin^2\left(k\rs\right)\, e^{-2k^2/\kD^2}
\nonumber
\\
%
&\approx 
9.4 a \int \frac{k \id k}{\kD^2}  \, \mathcal{P}_\zeta(k)\, 2\sin^2\left(k\rs\right)\, e^{-2k^2/\kD^2}
\end{align}
with $\mathcal{P}(k) \equiv k^3 P_\zeta(k)/2\pi^2$.
For a given $k$-mode, energy release happens when $k\simeq \kD(z)$, where $\kD$ is about $\sim 1.9\,(1+z)^{1/2}$ larger than the horizon scale $k_{\rm h}\approx \pot{2.1}{-6}(1+z)\,{\rm Mpc}^{-1}$, implying that small-scale power is dissipated well inside the horizon.
During the $\mu$-era, modes with $50\,{\rm Mpc}^{-1}\lesssim k \lesssim 10^4\,{\rm Mpc}^{-1}$ contribute most to the energy release, while $y$-distortions are mainly created by modes with $k\lesssim 50\,{\rm Mpc}^{-1}$.

\begin{figure}
\centering
\includegraphics[width=\columnwidth]{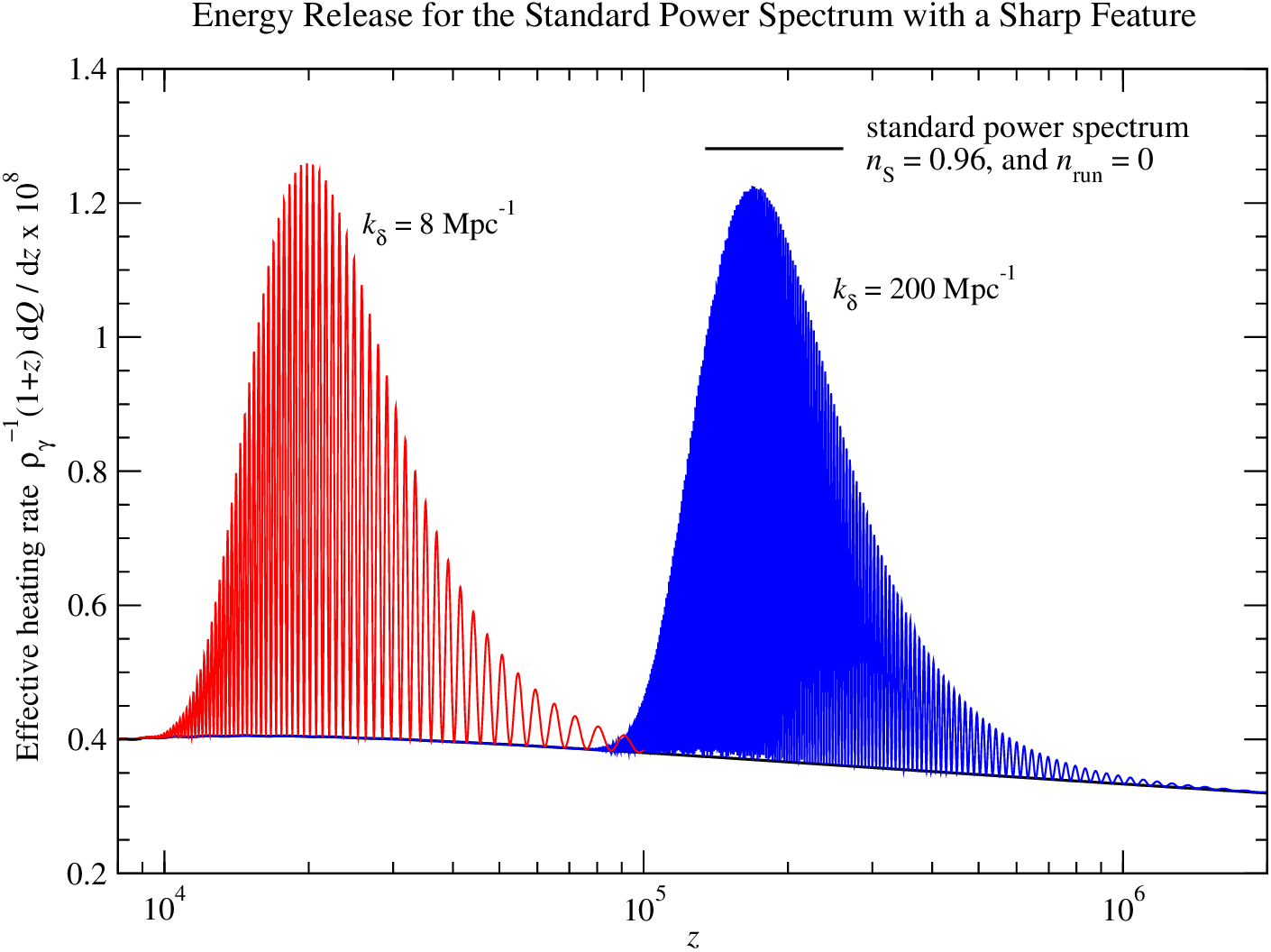}
\caption{Effective heating rate for the standard power spectrum, $P^{\rm st}_\zeta(k)$, with one sharp feature at $k_\delta$.
For illustration we chose $(\nS,\nrun)=(0.96,0)$. Furthermore, we set $A^\delta_\zeta=\pot{2.4}{-9}$ in both shown cases.
}
\label{fig:heating_rate_sharp}
\end{figure}
\subsection{Energy release by a single $k$-mode}
\label{sec:sharp}
We first consider the standard power spectrum with an extremely sharp feature at some scale $k_{\delta}$.
In this case the modification to the power spectrum is given by 
$\Delta P^\delta_\zeta(k)=2\pi^2 A^\delta_\zeta \, k^{-2}\delta(k-k_{\delta})$, where $A^\delta_\zeta>0$ determines the amplitude of the feature.
Inserting this into Eq.~\eqref{eq:Qdot_int} yields
\beal
\label{eq:heat_SZ_appr_sharp}
\frac{1}{\rho_\gamma}\frac{\id Q^\delta_{\rm ac}(z)}{\id z}
&\approx 
18.8 \,a \,A^\delta_\zeta\,[k_{\delta}/\kD]^2 \sin^2\left(k_{\delta}\rs\right)\, e^{-2k_{\delta}^2/\kD^2}
\end{align}
for the associated energy release. 
Notice that $\kD$, $\rs$, and $a$ are all functions of redshift.
This expression shows that power stored in a single $k$-mode is released over a rather wide range of redshifts. The energy release peaks close to 
\beal
\label{eq:z_diss}
z_{\rm diss}\approx \pot{4.5}{5}\left[\frac{k_{\delta}}{10^3\,{\rm Mpc^{-1}}}\right]^{2/3}
\end{align}
but oscillates rapidly {due to the sine part of the transfer function}. This is illustrated in Fig.~\ref{fig:heating_rate_sharp} for $k_\delta = 8\,{\rm Mpc^{-1}}$ and $k_\delta = 200\,{\rm Mpc^{-1}}$ with $A^\delta_\zeta=\pot{2.4}{-9}$ in both cases.
For $k_\delta = 8\,{\rm Mpc^{-1}}$ most of the energy is released during the $y$-era, while for $k_\delta = 200\,{\rm Mpc^{-1}}$ energy release occurs in the $\mu$-era.

Since the typical variation of the energy release rate is much longer than the oscillation period, one approximately has
\beal
\label{eq:heat_SZ_appr_sharp_av}
\frac{1}{\rho_\gamma}\frac{\id Q^\delta_{\rm ac}}{\id z}
&\approx 
9.4 \,a \,A^\delta_\zeta\,(k_{\delta}/\kD)^2 \, e^{-2k_{\delta}^2/\kD^2},
\end{align}
replacing $\sin^2(x)\rightarrow 1/2$, the average value over one oscillation.
The effect on the CMB spectrum can now be estimated by integrating the released energy over the redshifts relevant for the $\mu$-era and $y$-era. 
The $y$-era the integral can be performed analytically, while in the $\mu$-era effects related to the visibility of spectral distortions have to be taken into account, {i.e., see Eq.~\eqref{eq:def_mu_y_Dr_r}. With $\mu\approx 1.4 \,\Delta \rho_\gamma/\rho_\gamma|_\mu$ and $y\approx \frac{1}{4} \Delta \rho_\gamma/\rho_\gamma|_y$ the} corresponding estimates are well approximated by
\bsub
\label{eq:mu_y_sharp}
\beal
\label{eq:mu_y_sharp_a}
\mu_\delta
&\approx
2.2 \,A^\delta_\zeta 
\left[
 \exp\left(-\frac{\hat{k}_{\delta}}{5400}\right)
-\exp\left(-\left[\frac{\hat{k}_{\delta}}{31.6}\right]^2\right)
\right]
%
\\
\label{eq:mu_y_sharp_b}
y_\delta&
\approx 
%
0.4 \,A^\delta_\zeta 
\exp\left(-\left[\frac{\hat{k}_{\delta}}{31.6}\right]^2\right),
\end{align}
\esub
where $\hat{k}_{\delta}=k_{\delta}\,{\rm Mpc}$.
We mention here that the expression for $y$ is only expected to be valid for $k_\delta \gtrsim 1\,{\rm Mpc^{-1}}-5\,{\rm Mpc^{-1}}$. 
At larger scales, baryon loading, recombination effects, and free streaming become important \citep{Chluba2012}, all of which are neglected here.
These effects can be consistently treated using {\sc CosmoTherm}, but for the purpose of this work, the above expression provide useful estimates for the effect on the CMB spectrum over a wide range of $k$-values.
Notice also that for the $y$-type distortions we do not apply a sharp cutoff in redshift, but rather limit the range in $k$-space.

{\subsubsection{Distortion window-function in $k$-space}}
By replacing the amplitude $A^\delta_\zeta$ with $k^3 P_\zeta(k)/2\pi^2\equiv \mathcal{P}(k)$ and integrating over $\id \ln k$, it is possible to obtain estimates for the values of $\mu$ and $y$ for general primordial power spectra.
Since the expressions in Eq.~\eqref{eq:mu_y_sharp} are sufficiently simple, in many cases the integral over $\id \ln k$ even becomes analytic. For given small-scale power spectrum we find
\beal
\label{eq:mu_y_est_gen}
\mu
&\approx 2.2\!\int_{k_{\rm min}}^\infty 
\mathcal{P}_\zeta(k)
%
%
\left[
 \exp\left(-\frac{\hat{k}}{5400}\right)
-\exp\left(-\left[\frac{\hat{k}}{31.6}\right]^2\right)
\right]
\!\id \ln k
\nonumber
\\
y
&\approx 0.4 \!\int_{k_{\rm min}}^\infty 
\mathcal{P}_\zeta(k)
%
%
\exp\left(-\left[\frac{\hat{k}}{31.6}\right]^2\right) \id \ln k,
\end{align}
where we set ${k_{\rm min}}\simeq 1\,{\rm Mpc^{-1}}$
%
%
and $\hat{k}=k\,{\rm Mpc}$.
These expressions turn out to be very useful for estimates and simple computations, as we demonstrate below.
The exponential functions act as Green's function of the cosmological dissipation problem, and the expressions can be used for general power spectra, as long as the effect of dissipation at scales $k\lesssim 1\,{\rm Mpc^{-1}}$ is not important.
Modes in this range of wavenumbers are expected to affect the amplitude of the $y$-distortion, which has to be computed using a full perturbation calculation \citep{Chluba2012}.

{With these assumptions, Eq.~\eqref{eq:mu_y_est_gen} provides a weighted integral constraint on the small-scale power spectrum where $\mu_\delta$ and $y_\delta$ define {\it window functions} in $k$-space. For a given detection of $\mu$ this constraint has to be satisfied by any viable inflationary model. As we see below, COBE/FIRAS already placed interesting limits on several models. Furthermore, PIXIE will improve these limits by a large margin, strongly restricting possible inflaton trajectories.}
\\

\subsection{Energy release for the background power spectrum}
\label{sec:standard}
The total energy release and spectral distortions caused by the standard power spectrum, $P^{\rm st}_\zeta(k)$, according to Eq.~\eqref{eq:P_k_st}, were discussed in detail by \citet{Chluba2012}, with simple analytic approximations given for different values of $\nS$ and $\nrun$.
Here we are interested in cases with deviations from the standard shape occurring above some value of $k_{\rm crit}$.
Since the total $\mu$ and $y$-distortion are given by $\mu=\mu(k<k_{\rm crit})+\mu(k\geq k_{\rm crit})$ and $y=y(k<k_{\rm crit})+y(k\geq k_{\rm crit})$, to avoid double counting it is therefore useful to consider the {\it partial} energy release for the standard background power spectrum caused by modes with $k\geq k_{\rm crit}$. 
For many of our examples we shall assume $\nrun=0$. In this case one has \citep{Chluba2012}
%
\bsub
\label{eq:mu_y_ac_approx}
\beal
\label{eq:mu_ac_approx}
\mu^{\rm st}_{\rm ac}
&\approx \pot{5.54}{-4} A_\zeta \,\exp\left(9.92\,\nS^{1.23}\right),
\\
\label{eq:y_ac_approx}
y^{\rm st}_{\rm ac}
&\approx \pot{2.85}{-2} A_\zeta
\,\exp\left(4.32\,\nS^{1.53}\right),
\end{align}
\esub
for the {\it total} $\mu$ and $y$-parameters.
These expression were obtained using a detailed perturbation calculation carried out with {\sc CosmoTherm} \citep{Chluba2011therm}.

To compute the amount of energy release caused by modes with $k>k_{\rm crit}$ we start with the heating rate
\beal
\label{eq:heat_SZ_appr_std_nrun_0}
\frac{1}{\rho_\gamma}\frac{\id Q^{\rm st}_{\rm ac}}{\id z}
&\approx 
2.4 \,a \,A_\zeta\,\left(\frac{\kD}{\sqrt{2}\,k_0}\right)^{\nS-1} 
\!\!\Gamma\left(\frac{1+\nS}{2}, \frac{2k^2_{\rm crit}}{\kD^2}\right),
\end{align}
where $\Gamma(n,x)$ denotes the incomplete $\Gamma$-function.
For a scale-invariant power spectrum we can observe the redshift scaling $\rho_\gamma^{-1}\id Q^{\rm st}_{\rm ac}/\id z \propto 1/(1+z)$.
For this reason we usually present the effective heating rate as $(1+z)\,\rho_\gamma^{-1}\id Q^{\rm st}_{\rm ac}/\id z$.

Using Eq.~\eqref{eq:def_mu_y_Dr_r} one can easily compute the effective $\mu$ and $y$-parameters caused by energy release of modes $k>k_{\rm crit}$ numerically. 
Alternatively, with Eq.~\eqref{eq:mu_y_est_gen} we find
\beal
\label{eq:mu_y_std_k_kcrit}
\frac{\mu^{\rm st}_{\rm ac}(k\geq k_{\rm crit})}{2.2 A_\zeta}
&\approx 
\left(\frac{5400}{\hat{k}_0}\right)^{\nS-1}\!\!
\Gamma\left(\nS-1, \frac{\hat{k}_{\rm crit}}{5400}\right)
-
\frac{y^{\rm st}_{\rm ac}(k\geq k_{\rm crit})}{0.2\,A_\zeta}
,
\nonumber
\\
\frac{y^{\rm st}_{\rm ac}(k\geq k_{\rm crit})}{0.2 A_\zeta}
&\approx \left(\frac{31.6}{\hat{k}_0}\right)^{\nS-1}
\Gamma\left(\frac{\nS-1}{2}, \left[\frac{\hat{k}_{\rm crit}}{31.6}\right]^2\right),
\end{align}
where $\hat{k}\equiv k \, {\rm Mpc}$.
These expressions work very well for $\hat{k}_{\rm crit}\gtrsim 1$, however, for some examples we shall use the results obtained with a full perturbation calculation to derive constraints on parameters describing possible deviations for the standard power spectrum.

\section{Small-scale power spectrum constraints}
\label{sec:constraints}
In this section we discuss different small-scale power spectra, giving both the effective heating rates well before recombination, as well as possible constraints derived from $\mu$ and $y$-distortions.
%
%
For COBE/FIRAS the $2\sigma$ upper limits are $\mu \lesssim \pot{9}{-5}$ and $y \lesssim\pot{1.5}{-5}$ \citep{Fixsen1996}, while for PIXIE one expects $2\sigma$ detection limits of $\mu\simeq \pot{2}{-8}$ and $y \simeq\pot{4}{-9}$ \citep{Kogut2011PIXIE}.
When presenting results we usually assume these values, unless stated otherwise.

In the case of COBE/FIRAS this imposes a strong upper bound on the amplitude of the power spectrum, while for PIXIE the constraints should be interpreted as $2\sigma$-detection limits. Models above this limit should lead to a signal that can be detected at more than $2\sigma$ level, implying that they can be ruled out if no distortion is found.
We mention, however, that here we do not address the more difficult challenge of using the detection of a CMB spectral distortion to distinguish between different inflation scenarios.  
We furthermore take the optimistic point of view that foregrounds (e.g., due to synchrotron and free-free emission, dust and spinning dust) and systematics (e.g., frequency calibration, frequency-dependent beams) are sufficiently under control, so that the quoted detection limits of PIXIE can be truly achieved. More detailed forecasts including all these aspects will be required, but are beyond the scope of this paper.

\subsection{Instructive upper bounds on the amplitude of the power spectrum at small scales}
\label{sec:conservative}
Let us first consider the simplest ansatz for the small-scale power spectrum: assume that it is scale-independent with amplitude $\mathcal{P}_\zeta(k)= A_\zeta$ over some specified range of $k$.
If we estimate the $\mu$ and $y$-parameters for this case using Eq.~\eqref{eq:mu_y_est_gen}, and impose the COBE/FIRAS limits, we can determine an upper bound on $A_\zeta$. Clearly, the constraints on $A_\zeta$ weaken as the range of scales with enhanced power narrows.

\subsubsection{Optimistic upper limit on the amplitude of the small-scale power spectrum from COBE/FIRAS}
Since the power spectrum at large scales is well constrained by CMB anisotropies and LSS observations, we first assume that the small-scale power spectrum has a scale-independent amplitude [${\cal P}_\zeta(k) = 2.4\times10^{-9}$] at all scales with wavenumbers $k\lesssim 1\,{\rm Mpc}^{-1}$ and a different constant amplitude [${\cal P}_\zeta(k) = A_\zeta$] for $k\gtrsim 1\,{\rm Mpc}^{-1}$.
In this case, one can derive an upper limit on $A_\zeta$ from the $\mu$ and $y$-limits given by COBE/FIRAS; this is an optimistic constraint on the small-scale power spectrum because we have assumed that the power spectrum is equally enhanced on all scales that contribute to $\mu$ and $y$.

Carrying out the required integrals we find\footnote{We confirmed these results using CosmoTherm.} $\mu_{\rm op}\simeq 11\,A_\zeta$ and $y_{\rm op} \simeq 1.3 \,A_\zeta$, which implies $A^\mu_\zeta\lesssim \pot{8.4}{-6}$ and $A^y_\zeta\lesssim \pot{1.2}{-5}$.
Using the TRIS bound, $\mu\lesssim \pot{6}{-5}$ \citep{tris1, tris2}, one finds $A^\mu_\zeta\lesssim \pot{5.6}{-6}$.
According to the weight-functions defined in Eq.~\eqref{eq:mu_y_est_gen}, the $\mu$-limit is sensitive to power over $30\,{\rm Mpc}^{-1}\lesssim k\lesssim 10^4\,{\rm Mpc}^{-1}$, while the $y$-limit is driven by the $k$-range $1\,{\rm Mpc}^{-1}\lesssim k\lesssim 50\,{\rm Mpc}^{-1}$.
Therefore, these constraints on $A_\zeta$ are applicable to any power spectrum that has a constant amplitude over these ranges of scales.
(Note that for $y$ the lower cut-off is imposed by our assumptions, rather than for physical reasons.)

\subsubsection{Comparison to constraints from PBHs and UCMHs}
\label{sec:PBH_limits}
Another instructive example is motivated by the intention to compare the power spectrum constraints derived from CMB spectral distortions with those obtained from PBHs and UCMHs. 
To make this comparison, we must review how the latter are determined from observations 
that limit the abundance of PBHs and UCMHs.   
PBHs and UCMHs form in regions where the initial density contrast exceeds some critical value ($\delta\rho/\rho \gtrsim 0.3$ for PBHs and $\delta\rho/\rho \gtrsim 10^{-3}$ for UCMHs), and their masses are determined by the size of the overdense region that hosts them.  If the perturbations are assumed to be Gaussian, then the probability of forming a PBH or UCMH with a certain mass depends only on the mass variance within spheres that form PBHs and UCMHs with that mass.  Therefore, an upper limit on the abundance of PBHs and UCMHs with a given mass implies an upper bound on $\sigma_\mathrm{hor}^2(R)$: the density variance within a sphere of radius $R$ evaluated at horizon entry in total matter gauge \citep{JGM09, BSA11}.
These constraints on $\sigma_\mathrm{hor}^2(R)$ are then converted to constraints on the primordial curvature power spectrum ${\cal P}_\zeta(k)$ at wavenumber $k=R^{-1}$, but this conversion assumes a specific spectral shape for ${\cal P}_\zeta(k)$.

\begin{figure}
\centering
\includegraphics[width=\columnwidth]{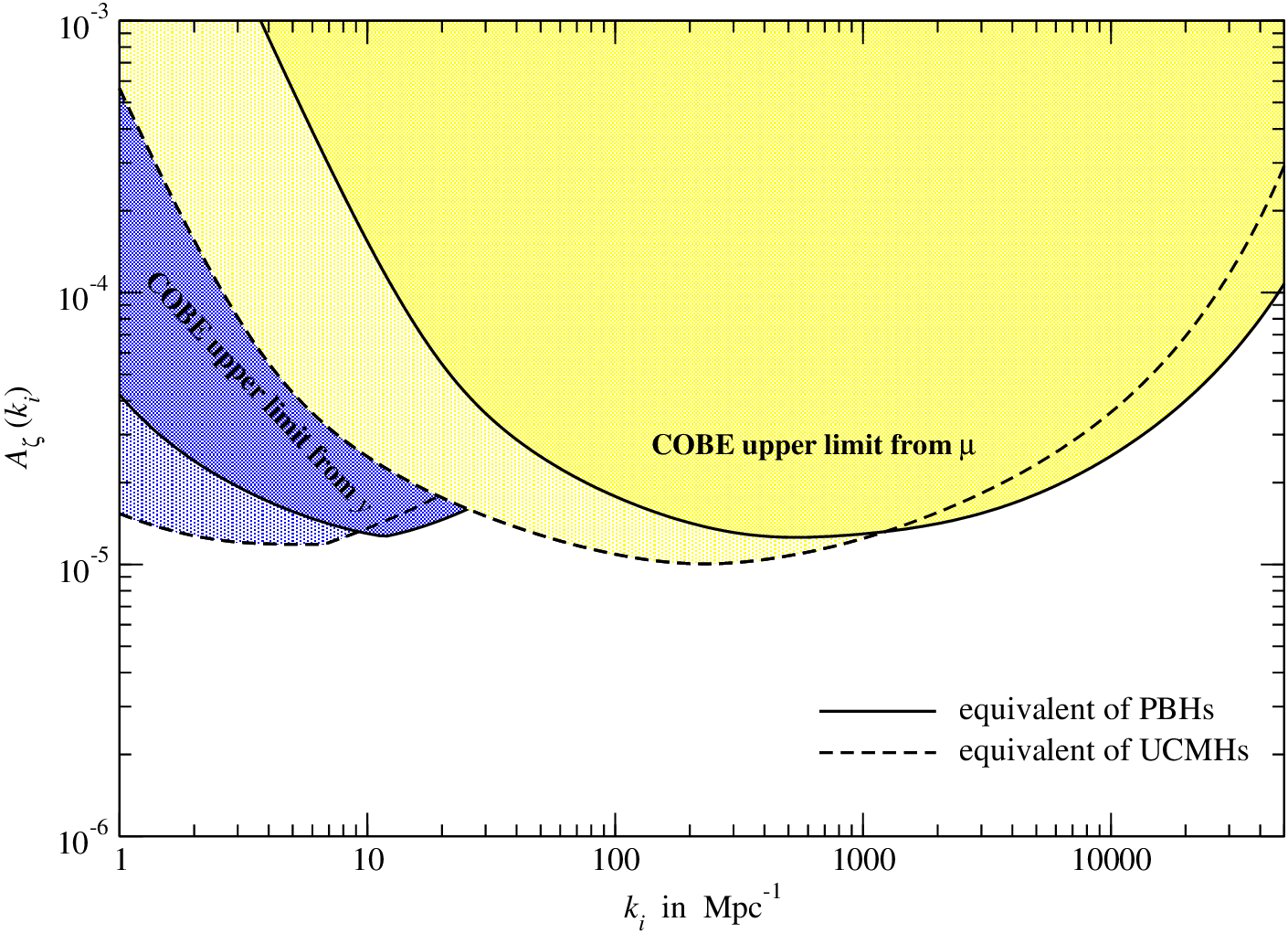}
\caption{
The bounds on $A_\zeta \equiv {\mathcal P}_\zeta$ derived from COBE/FIRAS with the same assumptions used to derive corresponding limits from PBHs and UCMHs.
}
\label{fig:constraint_PBHs}
\end{figure}
Since only the dark matter collapses to form a UCMH, the probability of UCMH formation depends on $\sigma_{\mathrm{hor},\chi}^2(R)$: the dark matter mass variance.  In contrast, the probability of forming a PBH depends on the total density perturbation at horizon entry, which is dominated by radiation.  Since dark matter perturbations and radiation perturbations evolve differently as they enter the horizon, $\sigma_\mathrm{hor}^2(R)$ and $\sigma_{\mathrm{hor},\chi}^2(R)$ have different definitions in terms of ${\cal P}_\zeta(k)$ \citep{JGM09, BSA11}.  Defining $x\equiv kR$, then
\bsub
\beal
\sigma_\mathrm{hor}^2(R) &= \frac{16}{3}\!\int_0^\infty \!\! x j^2_1\left(\frac{x}{\sqrt{3}}\right) F^2(x)\,{\cal P}_\zeta\left(k=\frac{x}{R}\right)\!\id x, \label{PBHsig}\\
\sigma_{\mathrm{hor},\chi}^2(R) &= \frac{1}{9}\!\int_0^\infty \!\!\! x^3 T_\chi^2\left(\theta=\frac{x}{\sqrt{3}}\right)  F^2(x)\, {\cal P}_\zeta\left(k=\frac{x}{R}\right) \!\id x \label{UCMHsig}
\end{align}
\esub
where $j_1(x)$ is a spherical Bessel function, $F(x)$ is a filter function, and 
\beq
T_\chi(\theta) \equiv \frac{6}{\theta^2}\left[\ln \theta +\gamma_{\rm E}-\frac{1}{2}-\mathrm{Ci}(\theta)+\frac{\sin \theta}{2\theta}\right],
\eeq
where $\gamma_{\rm E}$ is the Euler-Mascheroni constant and Ci is the cosine integral function.  When evaluating the constraints on ${\cal P}_\zeta(k)$ from PBHs, \citet{JGM09} use a Gaussian filter function, $F(x) = \exp(-x^2/2)$.  Meanwhile, \cite{BSA11} and \cite{LEW12} use the Fourier transform of a tophat window function, $F(x)= 3x^{-3}(\sin x - x \cos x)$, when evaluating the constraints on ${\cal P}_\zeta(k)$ from UCMHs.  In either case, if ${\cal P}_\zeta(k)$ is nearly scale-invariant, the integrals in Eqs. (\ref{PBHsig}) and (\ref{UCMHsig}) are dominated by the contribution from a narrow range of $x$ values around $x\simeq1$.   To derive constraints on ${\cal P}_\zeta(k)$ from the upper bounds on $\sigma_\mathrm{hor}^2(R)$ and $\sigma_{\mathrm{hor},\chi}^2(R)$ established by PBHs and UCMHs, it is customary to assume that ${\cal P}_\zeta(k)$ is \emph{locally} scale invariant, i.e. that it does not vary significantly over the limited range of scales that contribute to the mass variance at a given radius.  This assumption allows us to take ${\cal P}_\zeta(k)$ outside the integrals in Eqs. (\ref{PBHsig}) and (\ref{UCMHsig}), making $\sigma_\mathrm{hor}^2(R)$ and $\sigma_{\mathrm{hor},\chi}^2(R)$ proportional to ${\cal P}_\zeta\left(k=1/R\right)$.  

Since the $\mu$- and $y$- distortions produced by the dissipation of acoustic modes receive contributions from a much wider range of scales than the mass variance does, there is no model-independent way to compare the constraints on ${\cal P}_\zeta(k)$ from spectral distortions to those from PBHs and UCMHs.  Any such comparison requires one to specify the scale dependence of ${\cal P}_\zeta(k)$; we chose to make a comparison by applying the assumption of local scale invariance to the computation of CMB spectral distortions.  For each scale $k_i = 1/R$, we assume that ${\cal P}_\zeta(k)$ is nonzero only over the range of scales that contribute 99\% of the $\sigma_\mathrm{hor}^2(R)$ integral with a Gaussian filter ($0.085<k/k_i<2.485$) and 99\% of the $\sigma_\mathrm{hor,\chi}^2(R)$ integral with a tophat filter ($0.15<k/k_i<12.35$).  Within these scale ranges, we assume that ${\cal P}_\zeta(k) \equiv A_\zeta(k_i)$ and compute the resulting spectral distortions.   Since this power spectrum gives the same mass variance (to within 1\%) as a completely scale-invariant power spectrum, the constraints on the amplitude $A_\zeta(k_i)$ from COBE/FIRAS can now be directly compared to the constraints on the primordial power spectrum from PBHs and UCMHs; both sets of constraints make the same assumptions about the local scale-invariance of the power spectrum.

The results of our computation are summarized in Fig.~\ref{fig:constraint_PBHs}.
The typical limits for both the equivalent of the PBHs and UCMHs are $A_\zeta(k_i)\lesssim \pot{2}{-5}$ for $1\,{\rm Mpc}^{-1}\lesssim k_i\lesssim 10^4\,{\rm Mpc}^{-1}$. 
At smaller scales the bound becomes less stringent because the thermalization process starts being very efficient.
Notice also that for $k_i \lesssim 10\,{\rm Mpc}^{-1}$ the shape of the constraint derived from $y$ is affected by enforcing $k\gtrsim 1\,{\rm Mpc}^{-1}$. If we omit this restriction the curves become practically constant at a level $\simeq 10^{-5}$ for  $k_i \lesssim 10\,{\rm Mpc}^{-1}$; however, for these cases modification because of recombination, baryon loading and free streaming should be included to obtain accurate constraints.

For $10 \,\mathrm{Mpc}^{-1}\lesssim k \lesssim 10^4 \,\mathrm{Mpc}^{-1}$, the upper limits on $A_\zeta(k)$ from COBE/FIRAS are more than $10^3$ times stronger than the bound obtained from PBHs on these scales \citep[$A_\zeta \lesssim 0.06$;][]{JGM09}.  UCMHs can place stronger constraints on $A_\zeta(k)$ on these scales; if the mass of the dark matter dark matter particle is less than 5 TeV and it self-annihilates with $\langle \sigma v\rangle \geq 3\times 10^{-26} \, \mathrm{cm^3s^{-1}}$, then the fact that Fermi-LAT has not detected gamma-ray emission from UCMHS implies that $A_\zeta(k_i)\lesssim 2\times10^{-7}-3\times10^{-6}$ for $10 \,\mathrm{Mpc}^{-1}\lesssim k \lesssim 10^4 \,\mathrm{Mpc}^{-1}$ \citep{BSA11}.  Even if dark matter does not self-annihilate, UCMHs could still be used to slightly improve this bound; if Gaia does not detect microlensing by UCMHs, then $A_\zeta(k_i) \lesssim 10^{-5}$ for $k_i\simeq 3500\, \mathrm{Mpc}^{-1}$ for non-annihilating dark matter \citep{LEW12}.  
However, these constraints not only assume a particular density profile for UCMHs, but also that the UCMHs are not disrupted between their formation and today.
Therefore, the bounds on $A_\zeta(k)$ from COBE/FIRAS are more robust.
PIXIE could improve the bound on $A_\zeta(k)$ derived from measurement of the CMB spectrum by another two to three orders of magnitude, potentially reaching $A_\zeta(k_i)\lesssim \pot{\rm few}{-8}$ over scales $1 \,\mathrm{Mpc}^{-1}\lesssim k \lesssim 10^4 \,\mathrm{Mpc}^{-1}$.

\begin{figure}
\centering
\includegraphics[width=\columnwidth]{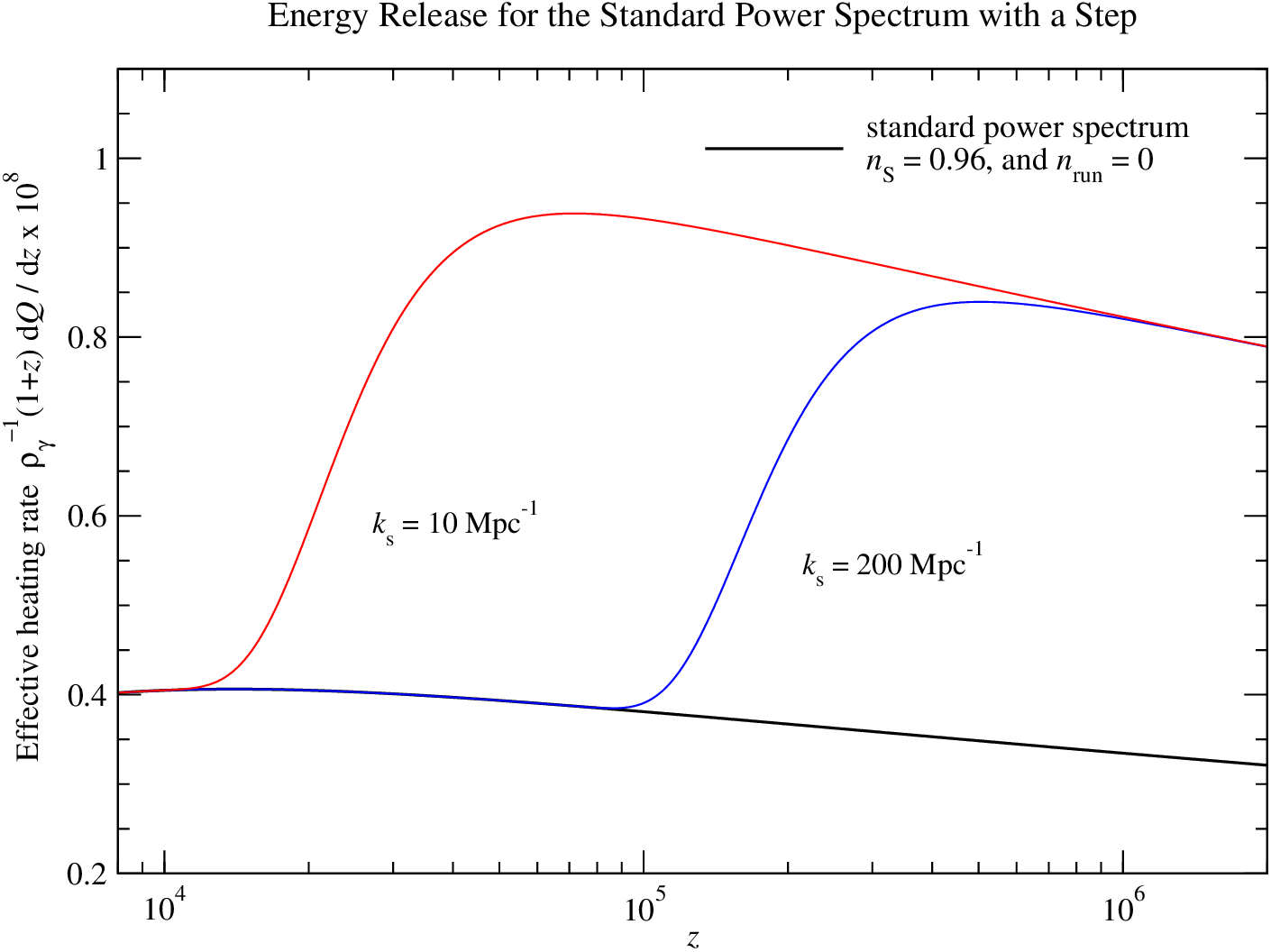}
\caption{Effective heating rate for the standard power spectrum, $P^{\rm st}_\zeta(k)$, with a step  at  $k_{\rm s}$.
For illustration we chose $(\nS,\nrun)=(0.96,0)$. Furthermore, we set $A^{\rm s}_\zeta=\pot{3.5}{-9}$ in both shown cases.
}
\label{fig:heating_rate_step}
\end{figure}
\begin{figure}
\centering
\includegraphics[width=\columnwidth]{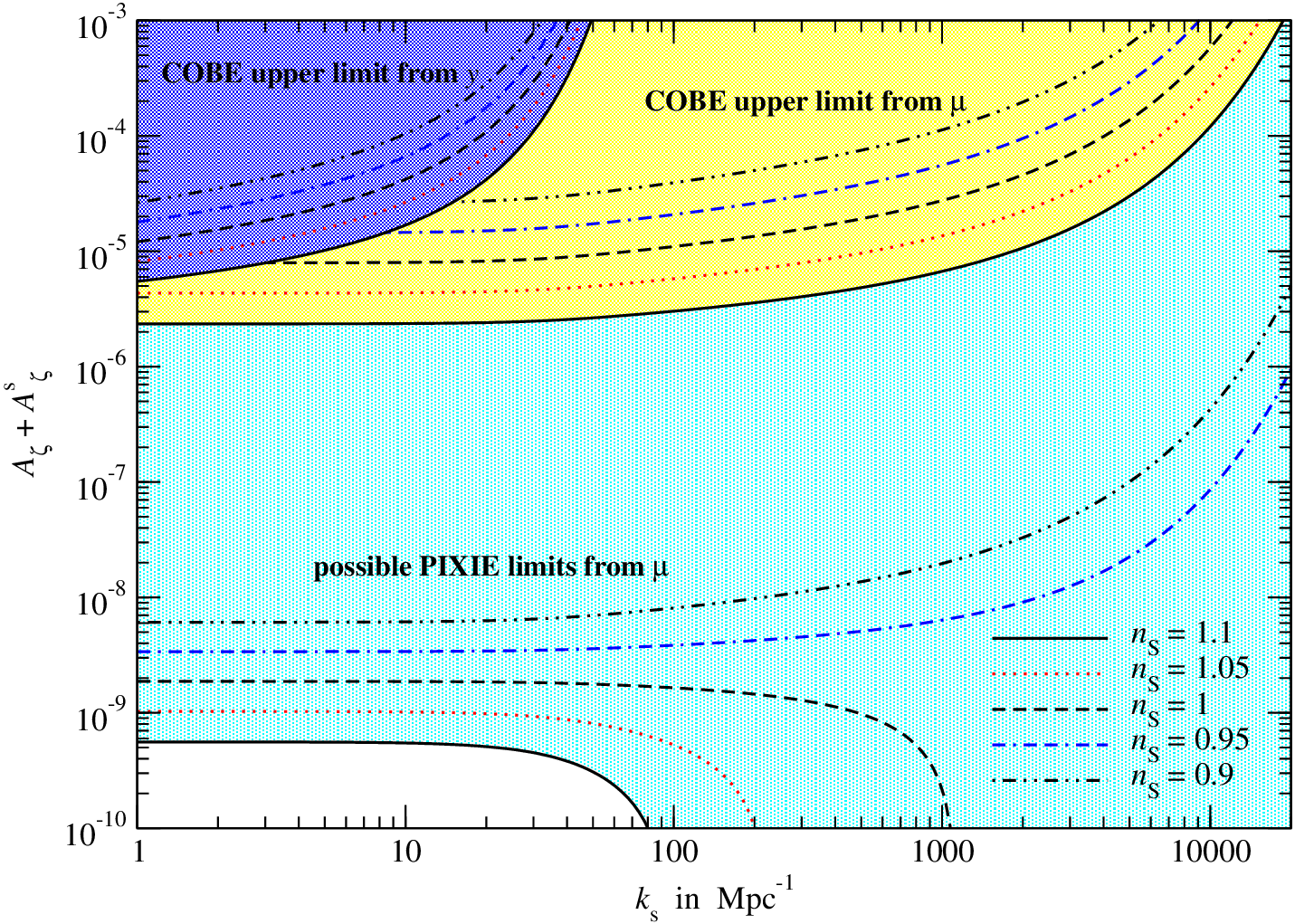}
\caption{Limits on the total amplitude of one step in the small-scale power spectrum occurring at $k_{\rm s}$. The energy release of modes with $k<k_{\rm s}$ was included, however, this only matters for the case of PIXIE. For illustration, the power law index was also varied.}
\label{fig:constraint_step}
\end{figure}
\subsection{Constraints on steps in the power spectrum}
\label{sec:steps}
Next, we consider a step in the power spectrum at some scale $k_{\rm s}$, where the amplitude changes from $A_\zeta$ for $k<k_{\rm s}$ to $A_\zeta+A^{\rm s}_\zeta$ for $k\geq k_{\rm s}$.
Such a step in the primordial power spectrum could be produced by multi-stage inflation models or inflaton potentials that change slope when the inflaton reaches a certain value \citep{1987PhRvD..35..419S, 1989PhRvD..40.1753S,1992JETPL..55..489S,Polarski:1992dq,1994PhRvD..50.7173I,1997NuPhB.503..405A,1998GrCo....4S..88S}.
We will assume that the spectral index is the same on both sides of the step, and we note that the step has to fulfill the condition $A^{\rm s}_\zeta\geq - A_\zeta$, as otherwise unphysical negative power is present in the power spectrum.
These power spectra can be parametrized by Eq.~(\ref{eq:P_k}) with $\Delta P_\zeta(k)=A^{\rm s}_\zeta\, P^{\rm st}_\zeta(k)/A_\zeta$ at $k\geq k_{\rm s}$ and $\Delta P_\zeta=0$ otherwise. 
If for simplicity we assume $\nrun=0$ for the background power spectrum, $P^{\rm st}_\zeta$, we find the effective heating rate by modes with $k\geq k_{\rm s}$ is simply given by Eq.~\eqref{eq:heat_SZ_appr_std_nrun_0} with $A_\zeta$ replaced by $A_\zeta+A^{\rm s}_\zeta$.
Similarly, the $\mu$ and $y$-parameters caused by the change in power can be estimated using the expressions Eq.~\eqref{eq:mu_y_std_k_kcrit}.

In Fig.~\ref{fig:heating_rate_step} we illustrate the time-dependence of the effective energy release 
for $k_{\rm s} = 10\,{\rm Mpc^{-1}}$ and $k_{\rm s} = 200\,{\rm Mpc^{-1}}$ with step amplitude $A^{\rm s}_\zeta=\pot{3.5}{-9}$ in both cases. The total amplitude of the power spectrum after the step therefore is $A^{\rm tot}_\zeta=A_\zeta+A^{\rm s}_\zeta\approx \pot{5.9}{-9}$.
In contrast to the single-mode case, we see that the energy release no longer exhibits any oscillatory behaviour, since oscillations of several neighbouring modes cancel each other, leading to smooth energy release.

In Fig.~\ref{fig:constraint_step} we present constraints on the total amplitude $A_\zeta+A^{\rm s}_\zeta$ of the power spectrum after the step.
We show both limits obtained from COBE/FIRAS and possible future bounds from PIXIE.
In the COBE/FIRAS case the energy release from the background spectrum, $P^{\rm st}_\zeta(k)$, can be neglected, as it only results in $\mu\simeq y\simeq 10^{-8}$ for reasonable values of $\nS$ and $\nrun$ \citep{Chluba2012}.
The COBE limits obtained from the $\mu$-distortion are most stringent in the range $k_{\rm s}\lesssim 10^3\,{\rm Mpc}^{-1}$; the limits get weaker at smaller scales because  the thermalization process starts being very efficient.
Also the $\mu$-limit is stronger than the $y$-limit because the $\mu$ distortion receives contributions from a slightly larger logarithmic range of wavenumbers ($1\,\mathrm{Mpc}^{-1}\lesssim {k}\lesssim 50 \,\mathrm{Mpc}^{-1}$ for $y$ as opposed to $50 \, \mathrm{Mpc}^{-1}\lesssim k\lesssim 10^4\,\mathrm{Mpc}^{-1}$ for $\mu$).

For PIXIE we only present the possible constraint derived by measurement of $\mu$. 
Obtaining a limit from $y$ is expected to be much harder, because at low redshifts many other astrophysical processes (e.g., energy release because of {\it supernovae} \citep{Oh2003}; shocks during large scale {\it structure formation} \citep{Sunyaev1972b, Cen1999, Miniati2000}; {\it unresolved SZ clusters} \citep{Markevitch1991}; the thermal SZ effect and second order Doppler effect from {\it reionization} \citep{McQuinn2005}) can cause an average $y$-distortion that is expected to be orders of magnitude larger in amplitude.
For PIXIE the energy release caused by the background power spectrum no longer can be ignored, since $\mu(k<k_{\rm s})\simeq \mu(k\geq k_{\rm s})$.
We therefore present the limit on just the amount of  dissipation at scales $k\geq k_{\rm s}$, after subtracting $\mu(k<k_{\rm s})$ as given by Eq.~\eqref{eq:mu_y_std_k_kcrit}.
For the shown examples with $\nS=1$, $\nS=1.05$ and $\nS=1.1$  {our results imply} that respectively for $k_{\rm s} \gtrsim 100\,{\rm Mpc^{-1}}$, $250\,{\rm Mpc^{-1}}$ and $10^3 \,{\rm Mpc^{-1}}$ a more than $2\sigma$-detection of $\mu$-distortions is expected even in the case $A_\zeta+A^{\rm s}_\zeta\equiv 0$ for $k>k_{\rm s}$.
This is simply because $\mu(k<k_{\rm s})$ itself already exceeds the $2\sigma$-detection limit of PIXIE, i.e., $\mu_{\rm lim}\simeq \pot{2}{-8}$.
Also, at about $k\simeq 20 \,{\rm Mpc^{-1}}$ the curves become flat, indicating the point at which practically no additional $\mu$-distortion is produced by modes with smaller wavenumber. In this regime only the amplitude of the $y$-distortion is expected to change; however, unless a large distortion ($y\gtrsim 10^{-6}$) is created, this signal will be hard to separate, as mentioned above.
Nevertheless, simultaneous detection of (large) $y$ and $\mu$ 
could constrain the scale at which the step occurred.

Here we only considered one step, but it is easy to generalize the discussion to multiple steps. 
The bound will strongly depend on the distribution of $k_{\rm s}$ and $A^{\rm s}_\zeta$ for which physical motivation should be provided, suggesting a case-by-case study is more useful.
If, for example, all steps increase the total power at small scales, then the derived limits are expected to become stronger.
However, for models with oscillations around a standard scale-invariant small-scale power spectrum \citep[see, e.g.,][]{2011JCAP...01..030A,2012arXiv1201.4848C}, the net effect should average out.

\subsection{Constraints on a bend in the power spectrum}
\label{sec:kink}
As a second example we consider a kink or bend in the power spectrum at some scale $k_{\rm b}$ with the slope of the power spectrum changing from $\nS$ to $\nS^\ast$, while $P(k)$ remains continuous. 
\citet{2008PhRvD..77b3514J} showed that such changes in the spectral index result from discontinuities in the second derivative of the inflaton potential, and this model for $P(k)$ may be used to approximate the power spectra produced by several other models that generate large perturbations on small scales \citep[e.g.,][]{Stewart1997, 2010JCAP...09..007B, 2011JCAP...03..028G, 2011JCAP...07..035L, 2011JCAP...11..028B, Shafi2011, Hotchkiss2012}. 
The associated power spectrum can be parametrized as 
\beal
\label{eq:P_bend}
P_\zeta(k)=
\begin{cases}
P^{\rm st}_\zeta (k) &\text{at} \, k< k_{\rm b} 
\\[1mm]
2\pi^2 k^{-3} \mathcal{P}^{\rm st}_\zeta(k_{\rm b})\,
(k/k_{\rm b})^{\nS^\ast-1} &\text{at} \, k\geq k_{\rm b}.
\end{cases}
\end{align}
with $\mathcal{P}^{\rm st}_\zeta(k)= k^3\,P^{\rm st}_\zeta(k)/2\pi^2$.
Assuming $\nrun=0$, the total power released by modes with $k>k_{\rm b}$ is again given by Eq.~\eqref{eq:heat_SZ_appr_std_nrun_0} with $A_\zeta$ replaced by\footnote{For Eq.~\eqref{eq:heat_SZ_appr_std_nrun_0} the pivot scale was $k_0$; to apply this expression one therefore has to rewrite $(k/k_{\rm b})^{\nS^\ast-1}=(k_0/k_{\rm b})^{\nS^\ast-1}(k/k_0)^{\nS^\ast-1}$, so that $P^{\rm st}_\zeta (k)=2\pi^2 k^{-3} \tilde{A}_\zeta \,(k/k_0)^{\nS^\ast-1}$ with $\tilde{A}_\zeta=\mathcal{P}^{\rm st}_\zeta(k_{\rm b})(k_0/k_{\rm b})^{\nS^\ast-1}=A_\zeta (k_{\rm b}/k_0)^{\nS-\nS^\ast}$ at wavenumbers $k\geq k_{\rm b}$.} 
$A_\zeta \, (k_{\rm b}/k_0)^{\nS-\nS^\ast}$.
Similarly, the $\mu$ and $y$-parameters caused by modes with $k>k_{\rm b}$ can be estimated using the expressions Eq.~\eqref{eq:mu_y_std_k_kcrit}.

\begin{figure}
\centering
\includegraphics[width=\columnwidth]{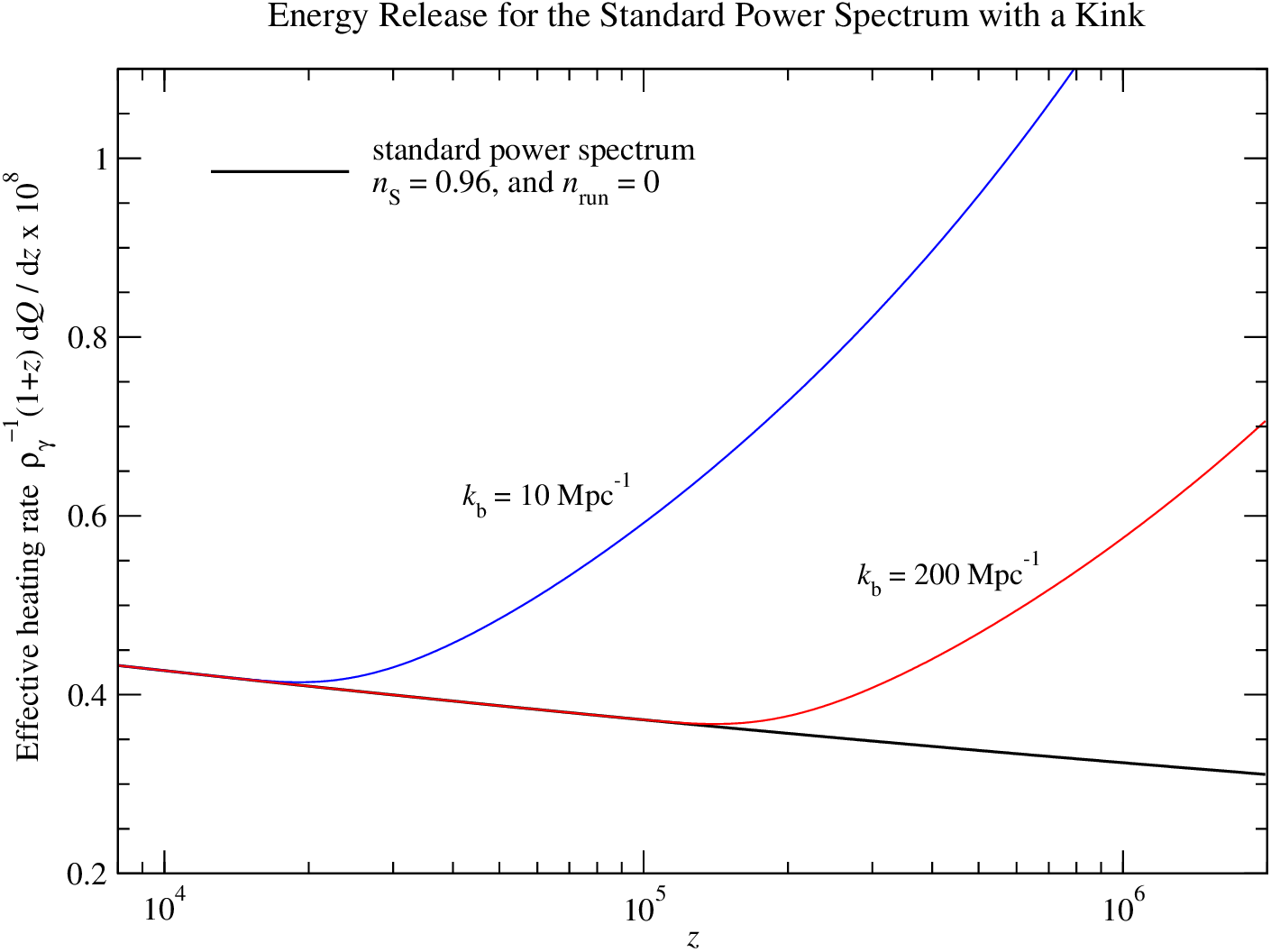}
\caption{Effective heating rate for the standard power spectrum with a kink at  $k_{\rm b}$.
For illustration we chose $(\nS,\nrun)=(0.96,0)$. Furthermore, we set $\nS^\ast=1.2$ in both examples.
}
\label{fig:heating_rate_kink}
\end{figure}
In Fig.~\ref{fig:heating_rate_kink} we illustrate the effective heating rate for $(\nS,\nrun)=(0.96,0)$ and $\nS^\ast=1.2$.
One can clearly see a flaring of the energy release that starts close to $z_{\rm diss}(k_{\rm b})$ according to Eq.~\eqref{eq:z_diss}.
The redshift dependence suggests that the effective $y$-parameter caused by 
a bend in the power spectrum is typically smaller than the $\mu$-parameter.
Indeed we find that for the COBE/FIRAS limits the constraints derived from $y$ are much weaker than those from $\mu$, so we neglect them for the discussion below.
Once again, concerns regarding confusion with low redshift $y$-distortions are the limiting factor for constraints derived from PIXIE's measurement of $y$, although simultaneous detection of both $\mu$ and $y$ from the cosmological dissipation process would provide a deeper understanding of the shape of the small-scale power spectrum and the position of a possible kink.

\begin{figure}
\centering
\includegraphics[width=\columnwidth]{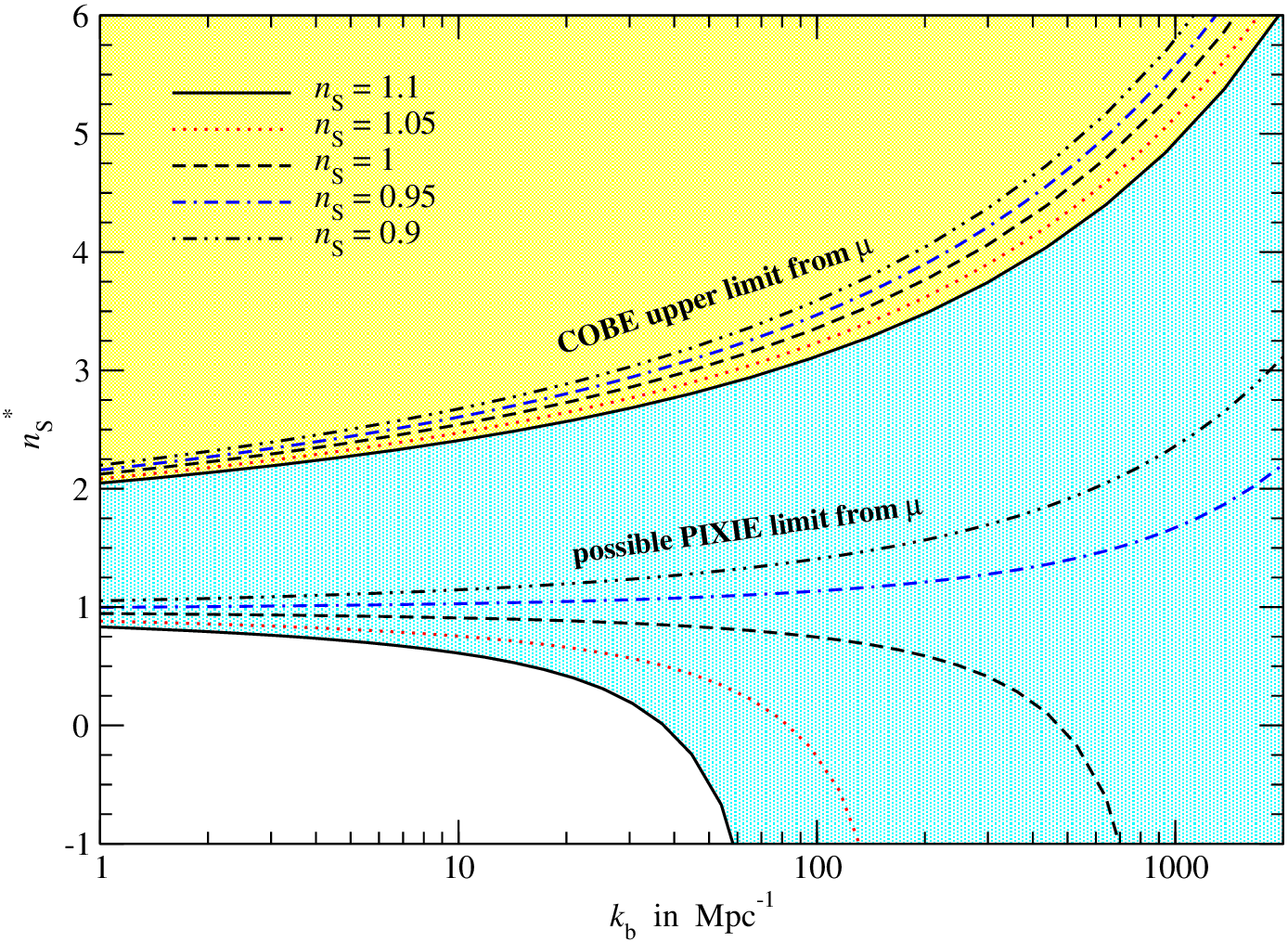}
\caption{Limits on the power law index $\nS^\ast$ for a bend in the small-scale power spectrum at $k_{\rm b}$. We assumed that the background power spectrum has no running. The limits obtained from $y$-distortions were always much weaker and hence have been omitted.}
\label{fig:constraint_kink}
\end{figure}

We can derive constraints on the value of $\nS^\ast$, and these constraints are shown in Fig.~\ref{fig:constraint_kink}.
The COBE/FIRAS limit on $\mu$ already rules out changes in the power law index by $\Delta \nS\equiv \nS^\ast-\nS \gtrsim 1$ at $k_{\rm b}\simeq 1\,{\rm Mpc^{-1}}$ at $2\sigma$-level.
With PIXIE this measurement will be strongly improved. For example, if the background spectrum has $\nS\simeq 0.96$ then even at $k_{\rm b} \simeq 100\,{\rm Mpc^{-1}}$ the slope cannot change by more than $\Delta \nS\simeq 0.1$ without leading to an observable $\mu$-distortion. 
Also, like in the step case, for $\nS\gtrsim 1$ a bend at large values of $k_{\rm b}$ should lead to an observable signal even if the power spectrum cuts off abruptly  (i.e., $\nS^\ast$ has a large negative value).
%
This should allow placing very tight constraints on inflationary models with flaring power spectra at small scales.

We note that PIXIE cannot constrain hybrid inflation models that use a `waterfall' field to end inflation \citep{2011JCAP...03..028G, 2011JCAP...07..035L,2011JCAP...11..028B} because these models predict $k_{\rm b} \gg 10^4$ Mpc$^{-1}$, corresponding to scales that left the horizon during the last few e-folds of inflation. Even PIXIE will be insensitive to energy release from those scales.

\begin{figure}
\centering
\includegraphics[width=\columnwidth]{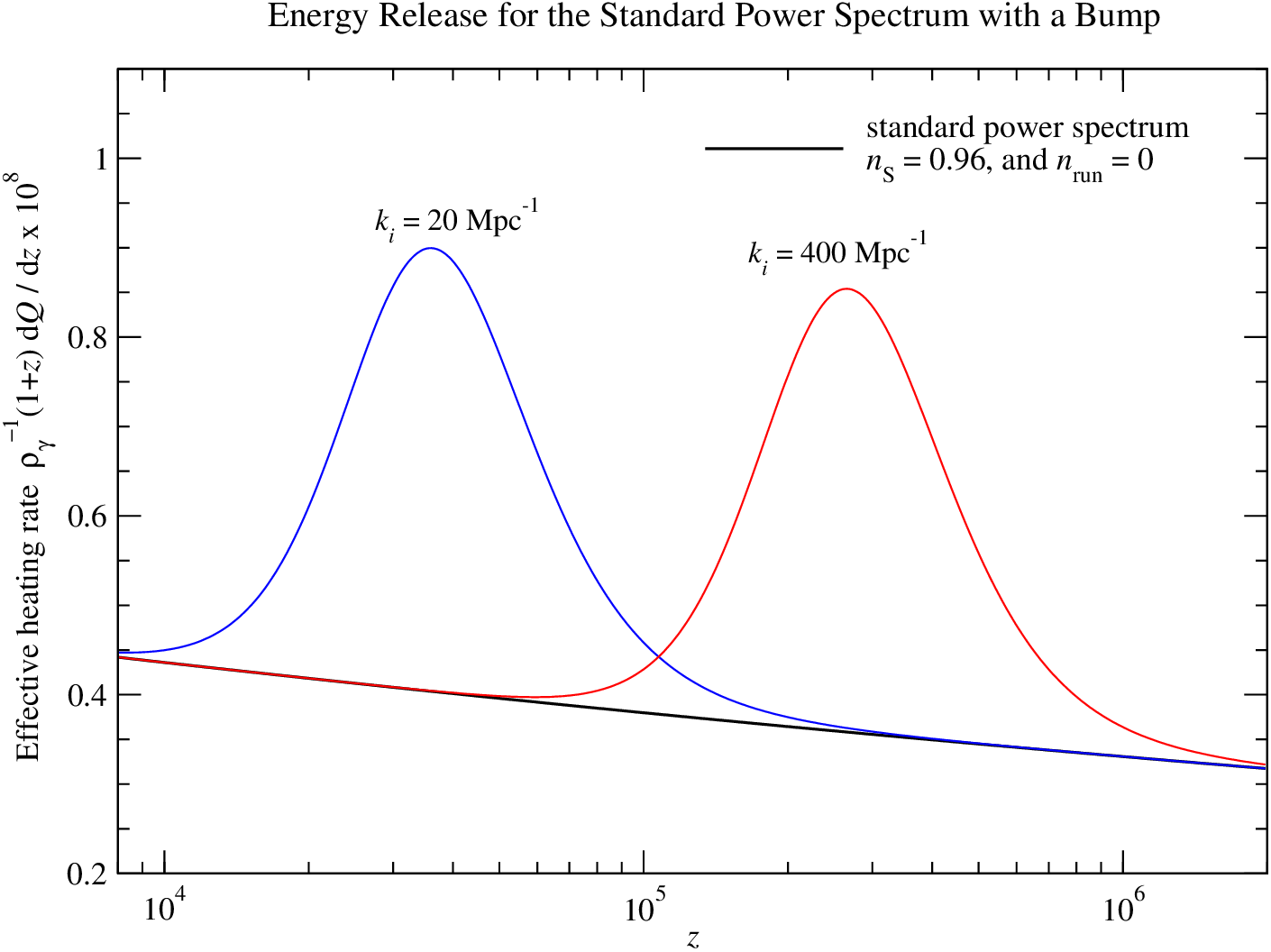}
\caption{Effective heating rate for the standard power spectrum with a feature at $k_i$ caused by particle production during inflation.
For illustration we chose $(\nS,\nrun)=(0.96,0)$ for the background spectrum. Furthermore, we set $A^{{\rm p},i}_\zeta=\pot{3.5}{-9}$ in both shown cases.
}
\label{fig:heating_rate_particles}
\end{figure}

\subsection{Constraints on particle production during inflation}
\label{sec:Neil}
\citet{Neil2009I} and \citet{Barnaby2010} showed that bursts of particle production during inflation produce localized bumps in the primordial power spectrum.  Specifically, if the inflaton $\phi$ is coupled to another scalar field $\chi$ via an interaction given by %
${\cal L}_\mathrm{int} = -\frac{1}{2}g_i^2(\phi-\phi_i)^2\chi^2$,
then the $\chi$ particles are temporarily massless when $\phi=\phi_i$.  At this time, $\chi$ particles are created by quantum effects, and these particles quickly become massive as $\phi$ moves away from $\phi_i$.  The massive $\chi$ particles then rescatter off the $\phi$ field, generating perturbations in $\phi$ that freeze once their wavelength exceeds the Hubble distance.  The massive $\chi$ particles are rapidly diluted by the inflationary expansion, so only a limited range of scales receive extra perturbations.

\citet{Neil2009} provided a simple parametrization for the the resulting bump in the primordial power spectrum: 
\beal
\label{eq:P_k_Neil}
\Delta P^{{\rm p},i}_\zeta(k)&=2\pi^2 A^{{\rm p},i}_\zeta \left(\frac{\pi e}{3k_i^2}\right)^{3/2} \exp\left(-\frac{\pi}{2}\frac{k^2}{k^2_i}\right).
\end{align}
The amplitude of the feature, $A^{{\rm p},i}_\zeta$, is simply related to the value of coupling constant $g_i$;  $A^{{\rm p},i}_\zeta\simeq \pot{1.01}{-6} g^{15/4}_i$.  The derivation of this feature in the primordial power spectrum is only valid for $10^{-7}\lesssim g_i^2\lesssim 1$ \citep{Neil2009I}, so we are primarily interested in $A^{{\rm p},i}_\zeta$ values between $10^{-19}$ and $10^{-6}$.
In contrast, there are no restrictions on the location of the bump; $k_i$ is determined by the number of $e$-foldings between the moment when $\phi=\phi_i$ and the end of inflation.
There may also be other fields with the same coupling to the inflation, each with their own values for $g_i$ and $\phi_i$.  In this case, the power spectrum will contain multiple bumps, and one should sum the contributions from each episode of particle production.

Inserting Eq.~\ref{eq:P_k_Neil} into Eq.~\eqref{eq:Qdot_int}, 
we find
\beal
\label{eq:Qdot_Neil}
\frac{1}{\rho_\gamma}\frac{\id Q_{{\rm p},i}}{\id z}
&\approx 
9.4 a\,A^{{\rm p},i}_\zeta \frac{e^{3/2}}{\sqrt{6\pi}}\,\frac{(k_i/\kD)^2}{\left[1+\frac{4}{\pi}(k_i/\kD)^2\right]^{5/2}}
\end{align}
for the effective heating rate of one feature at high redshifts {(see Fig.~\ref{fig:heating_rate_particles} for illustration)}.
The $\mu$ and $y$-parameter caused by one feature are roughly given by 
\bsub
\label{eq:mu_y_part}
\beal
\label{eq:mu_y_part_a}
\mu_{\rm p}
&\approx
2.4 \,A^{{\rm p},i}_\zeta\,
\left(e^{\xi^2_i}(1+2\xi^2_i)\,{\rm Erfc}(\xi_i)-\frac{2\xi_i}{\sqrt{\pi}}
-\frac{y_{\rm p}}{0.4\,A^{{\rm p},i}_\zeta}\right)
\\
\label{eq:mu_y_part_b}
y_{\rm p}&
\approx 
0.4 \,A^{{\rm p},i}_\zeta\,\left[1+(\hat{k}_i/40)^2\right]^{-3/2},
\end{align}
\esub
where $\hat{k}_i=k_i\,{\rm Mpc}$, $\xi_i=\hat{k}_i/\pot{1.35}{4}$, and ${\rm Erfc}(x)$ is the complementary error function. 
We again made use of Eq.~\eqref{eq:mu_y_est_gen} to give these simple expressions, but we also confirmed the validity of these expressions by numerically evaluating the nested integrals of the power spectrum and the heating rate.
%

\begin{figure}
\centering
\includegraphics[width=\columnwidth]{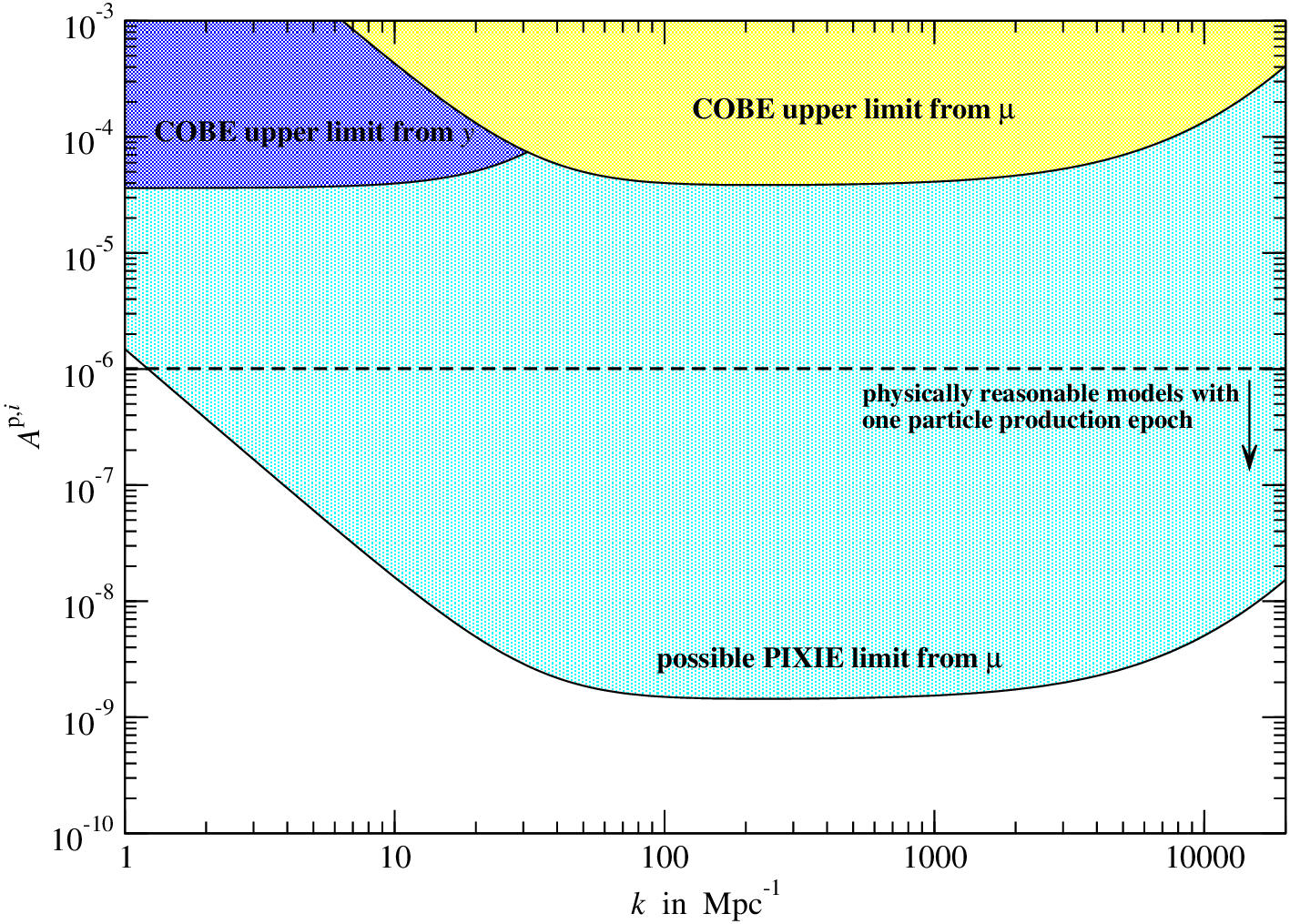}
\caption{Limits on inflation models with particle production. Here we only considered one episode of particle production. The dashed line indicates the bound $A^{{\rm p},i}_\zeta\lesssim10^{-6}$, required to ensure physically reasonable models \citep{Neil2009}. We assumed a background power spectrum with $(\nS,\nrun)=(0.96,0)$.}
\label{fig:constraint_particles}
\end{figure}
In Fig.~\ref{fig:constraint_particles} we present the derived limits on the amplitude for one episode of particle production.
The limits derived from COBE/FIRAS are weaker than the bound $A^{{\rm p},i}_\zeta\lesssim10^{-6}$ that is required to make the underlying calculation self-consistent. 
However, these bounds are still interesting, as they can be also interpreted as rather tight constraints on any other inflation models with bump-like features in the small-scale power spectrum that have a typical total width of $\Delta k/k\sim 2/\sqrt{3}$.

The bound becomes much tighter for PIXIE, basically limiting $A^{{\rm p},i}_\zeta\lesssim \pot{2}{-9}$ in the range $10^2\,{\rm Mpc^{-1}}\lesssim k\lesssim 10^3\,{\rm Mpc^{-1}}$.
Also, in the range $1\,{\rm Mpc^{-1}}\lesssim k \lesssim 100\,{\rm Mpc^{-1}}$, the limit derived from $\mu$ alone is still very interesting, although it is weaker since fewer modes related to the bump are able to release energy in the $\mu$-era.
Similarly, the bound becomes less stringent for $k\gtrsim 10^3\,{\rm Mpc^{-1}}$ because thermalization becomes efficient. 
Features in the small-scale power spectrum introduced by particle production are rather broad, with a significant tail of energy release towards lower redshifts, where the visibility for spectral distortions increases. Therefore, even for larger $k_i$ an observable $\mu$-distortions is created, softening the thermalization cut-off.
This implies that PIXIE could even constrain episodes of particle production with $k_i$ up to $\sim \pot{5}{4}\,{\rm Mpc^{-1}}$ in an interesting way, complementing the limits obtained from  CMB anisotropies and LSS at larger scales \citep{Neil2009}.

We also mention that for more than one episode of particle production, the constraints should become tighter. However, in this case again physical motivation should be given and models ought to be discussed on a case-by-case basis.
Furthermore, the constraint on bumps in the power spectrum also depends on the background power spectrum at small scales. 
For models with $\nS\simeq 1$ direct detection of the $\mu$-distortion  {with PIXIE could be possible even without extra bumps}, so that any excess power added by particle production features should further enhance the $\mu$-distortion  {above the $2\sigma$ detection threshold.
This indicates that the interpretation of the constraint depends significantly on assumptions about the background model extrapolated from CMB and LSS scales all the way to ${k}\simeq 10^4\,\mathrm{Mpc}^{-1}$.

\begin{figure}
\centering
\includegraphics[width=\columnwidth]{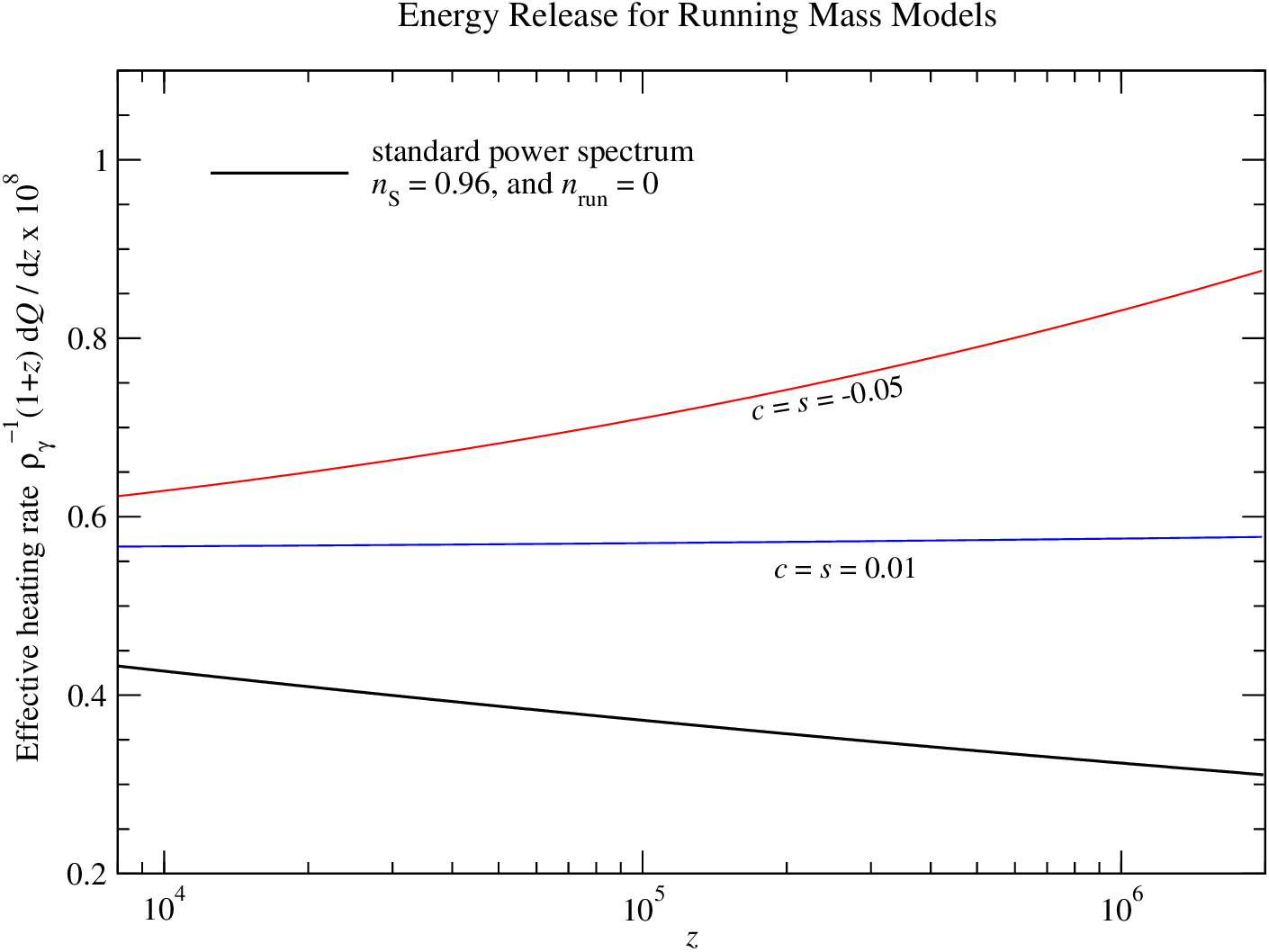}
\caption{Effective heating rate for inflation models with running-mass.
The power spectrum is parametrized as in Eq.~\eqref{eq:running_Pk}.
For reference we also show the standard power spectrum for $(\nS,\nrun)=(0.96,0)$.
}
\label{fig:heating_rate_running_mass}
\end{figure}
\subsection{Constraining running-mass inflation models}
\label{sec:running_mass}
For our next explicit example, we consider  running-mass inflation, which is a single-field supersymmetric inflation model that can generate enhanced power on small scales \citep{Stewart1997, Stewart1997b, 1999PhRvD..59f3515C, 1999PhRvD..60b3509C}.  These models assume that the dominant loop correction to the inflaton potential is $\propto\phi^2 \ln[\phi/Q]$, where $Q$ is the renormalization scale.  In this case, the renormalization-group-improved inflaton potential is $V(\phi) = V_0 + m^2(\ln\phi)\phi^2/2$, implying that the mass of the inflaton effectively changes during inflation. 
Consequently, running-mass models offer a solution to the $\eta$-problem of supergravity inflation; in supergravity, scalar fields usually have masses that are too large to drive inflation, but in running-mass inflation, $|m^2|$ can be large when $\phi$ is equal to the reduced Planck mass $M_\mathrm{Pl}$ while being small enough to permit inflation at smaller values of $\phi$.

To ensure that the inflaton potential is sufficiently flat, inflation must occur near an extremum of the inflaton potential.  Therefore, we can approximate the inflaton mass as
\beq
m^2 = \frac{c}{2} \frac{V_0}{M_\mathrm{pl}^2}\left[1 - 2\ln \frac{\phi}{\phi_*}\right],
\nonumber
\eeq
where $\phi_*$ is the value of $\phi$ at the extremum of $V(\phi)$ and $c \equiv -(M_\mathrm{Pl}^2/V_0) \id m^2/\id \ln\phi$ evaluated at $\phi=\phi_*$.  The $c$-parameter may be positive or negative, but explicit formulations of running-mass inflation in the context of supersymmetry have $0<c<1$ \citep[e.g.][]{Covi2004}.}  Given this linear approximation for $m^2(\ln \phi)$, 
the primordial power spectrum can be parametrized as \citep[cf.][]{1999PhRvD..59f3515C,Covi2004} 
\beq
\label{eq:running_Pk}
{\mathcal P}(k) = {\mathcal P}(k_0)\exp\left[\frac{2s}{c}\left(e^{c\Delta N(k)}-1\right)-2c\Delta N(k)\right].
\eeq
In this expression $\Delta N(k) \equiv \ln(k/k_0)$, and $s=c \ln(\phi_*/\phi_0)$, where $\phi_0$ is the value of the inflaton when the mode with wavenumber $k_0$ exited the Hubble horizon, $N_0$ e-folds before the end of inflation.  Note that $s$ can be positive or negative, depending on the sign of $c$ and the direction $\phi$ rolls during inflation.  
It follows that $s = c \ln(\phi_*/\phi_{\rm e}) \exp[-cN_0]$, where $\phi_{\rm e}$ is the inflaton value at the end of inflation.  Therefore, values of $s$ with $|s| \lesssim |c| e^{-cN_0}$ imply that $\phi_{\rm e}\simeq \phi_*$, which requires certain fine-tuning.

\begin{figure}
\centering
\includegraphics[width=0.85\columnwidth]{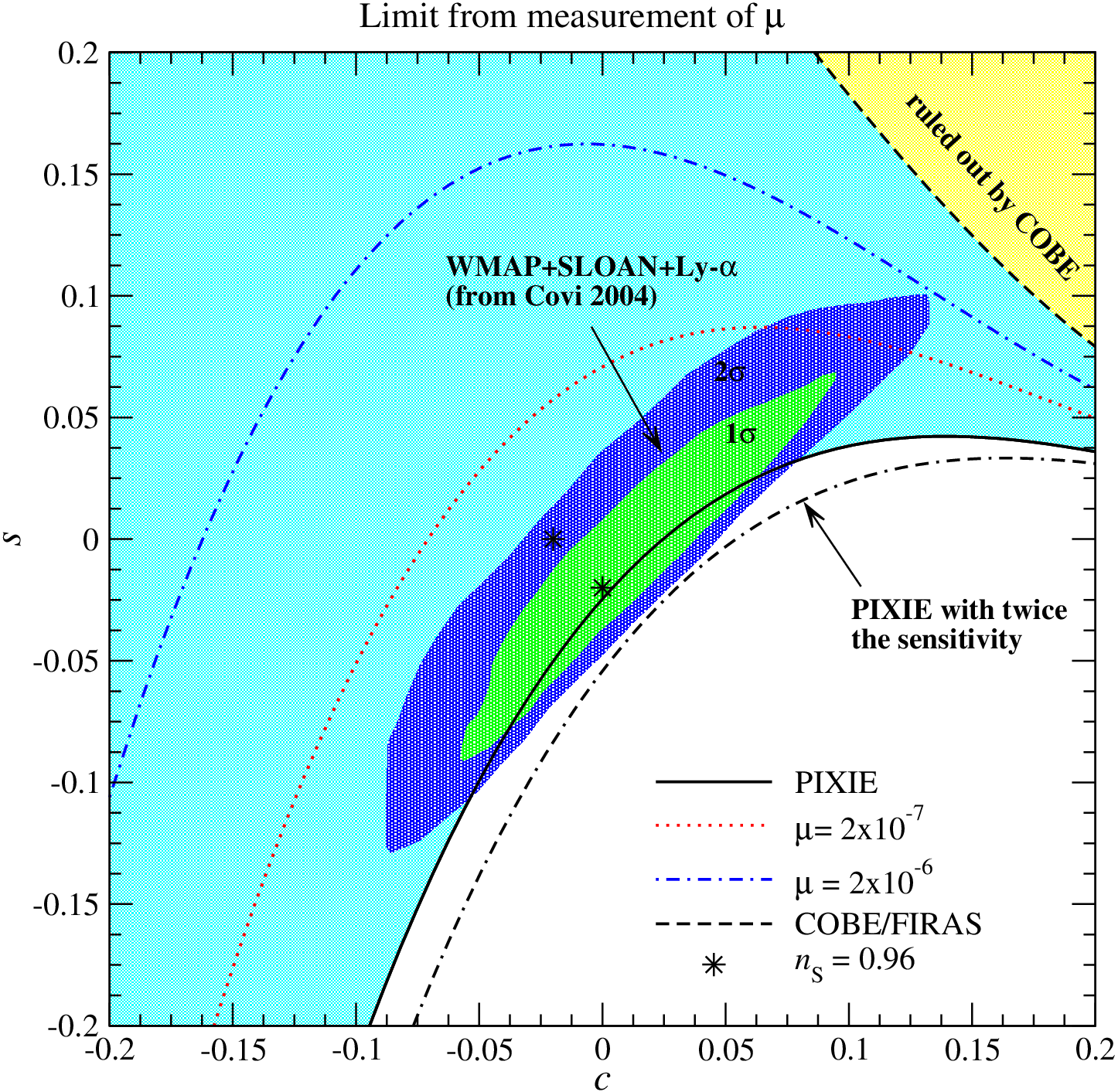}
\caption{Constraints on running-mass inflation models obtain by measurements of $\mu$. The closed contours show the constraints obtained by \citep{Covi2004}. PIXIE might help narrowing down the allowed parameter space to a small region around $s\simeq c-0.033$ with $-0.03\lesssim c \lesssim 0.06$. Furthermore, PIXIE with twice the sensitivity could in principle rule out running-mass inflation models  in case no distortion is found. For comparison we also show the location of the {\it two} models that are consistent with $(\nS,\nrun)=(0.96,0)$ at the pivot scale $k_0$.}
\label{fig:constraint_running_mass}
\end{figure}

One can also directly relate $s$ and $c$ to the usual  spectral index and running \citep{Covi2004}:
\bsub
\label{eq:nS_nrun_running}
\beal
\nS& \approx1+2\left[s (1-1.06 c)-c\right]+ \frac{2}{3}\left(s -c\right)^2,
\\
\nrun&\approx 2 s c.
\end{align}
\esub
Observational parameter limits on $s$ and $c$ derived using previous WMAP \citep{WMAP_params}, SDSS \citep{Tegmark2004} and Ly-$\alpha$ forest \citep{Seljak2005} measurements are shown as closed contours in Fig.~\ref{fig:constraint_running_mass} \citep{Covi2004}, limiting the range of allowed models to cases with $s \simeq c$ and $|s,c|\lesssim 0.13$.
According to Eq.~\eqref{eq:nS_nrun_running}, the allowed models typically have positive running with $-0.001 \lesssim \nrun\lesssim 0.02$ and $0.91\lesssim \nS\lesssim 1.06$.
This indicates that updated constraints from the latest CMB and LSS measurements that favor negative running might already further narrow down the allowed parameter space in comparison to \citet{Covi2004}.
However, the two running-mass models that match the current best-fit WMAP7 model without running, $(\nS,\nrun)=(0.96,0)$ at $k=k_0$ are still viable (cf. Fig.~\ref{fig:constraint_running_mass}).

Using Eq.~\eqref{eq:running_Pk} we can easily compute the effective heating rate and distortion parameters for different values of $s$ and $c$.
In Fig.~\ref{fig:heating_rate_running_mass} we illustrate this for two running-mass models that are in agreement with the constraints of \citet{Covi2004}.
The departure from the standard background spectrum occurs very gradually, so that in comparison with the standard power spectrum, both enhanced $\mu$ and $y$-distortions are expected in nearly all cases.

To compute the limits on the parameters $s$ and $c$ shown in Fig.~\ref{fig:constraint_running_mass}, we use the expressions given by Eq.~\eqref{eq:mu_y_est_gen}.  We also confirmed that a more precise computation practically gives the same result.
The COBE/FIRAS limit on $\mu$ does rule out a significant part of the theoretically allowed parameter space, however, in comparison to the constraints given by \citet{Covi2004} no improvement is achieved. 
On the other hand, PIXIE might rule out many running-mass inflation models at the $2\sigma$ level, narrowing the possible parameter space down to a slim region around $s\simeq c-0.033$ and $-0.03\lesssim c \lesssim 0.06$.
Furthermore, PIXIE with twice the sensitivity could rule out running-mass inflation models if no distortion is detected.
We also computed the limit from measurement of the $y$-parameter, but found that the constraint was always much weaker.

\subsection{Spectral distortions for small-field models generating detectable gravitational waves}
\label{sec:Ido}
%
Another interesting class of inflation models that predicts an enhancement of the power spectrum at small scales is a class of small-field models that generate a detectable gravitational wave (GW) signal \citep{2010JCAP...09..007B, Shafi2011, Hotchkiss2012}. 
Within this class, models with larger GW signal are also expected to produce excess power at small scales. Therefore, the detection of both a GW signal and a CMB spectral distortion at the predicted level would provide strong evidence for these models.

The exact power spectrum has to be evaluated numerically \citep[as done in][]{2010JCAP...09..007B} and exhibits a richer phenomenology than the standard nearly scale-invariant case, even when $\nrun$ is included. 
The ability to produce a detectable GW signal with limited field excursion, $\Delta \phi \lesssim M_{\rm Pl}$, is based on the fact that the slow-roll parameter $\epsilon$ is non-monotonic from the time the CMB scales left the horizon until inflation ends.
More explicitly, $\epsilon$ is initially large enough to produce a detectable GW signal, then it decreases to a smaller value for most of the duration of inflation.  At a later time, $\epsilon$ increases to unity and inflation ends. 
The scales of $10^{-3}\,{\rm Mpc^{-1}}<k<10^4\,{\rm Mpc^{-1}}$ exit the horizon while $\epsilon$ is decreasing;  since $\mathcal{P}_{\zeta} \propto V/\epsilon$, small-scale power is enhanced. 
The power spectrum also exhibits a maximum at a smaller scale that exits the horizon just before $\epsilon$ begins to increase.

Since here we are only interested in the possible CMB distortions we do not repeat the exact numerical calculation, but use the `improved power spectrum' form \citep[Eq.~(78) of][]{Lyth1999}, which captures all essential features of the model.
A more complete likelihood analysis is deferred to later work.
We calculated the $\mu$ and $y$ distortions for some models discussed in \citet{2010JCAP...09..007B}, however, we added specific models\footnote{Following the definitions of \citet{2010JCAP...09..007B}, these correspond to: $\alpha_0=\eta_0=0$ and $\phi_{\rm END}=1$; $\alpha_0=0$, $\eta_0=-0.02$ and $\phi_{\rm END}=1$.} with small tensor to scalar ratio $r\simeq 10^{-3}$. The GW signal of these scenarios could potentially be detected with PIXIE.

As mentioned above, we find that the level of $\mu$ and $y$ distortions is linked to the GW signal. This is because a larger GW signal requires a larger change in $\epsilon$ between CMB scales and smaller scales. We therefore expect a bigger enhancement of small-scale perturbations. For one of the models with relatively large GW signal, $r=0.08$, we obtain $\mu\simeq 8.5 \times 10^{-6}$ and $y\simeq 1.3 \times 10^{-7}$.  At this level, we expect a definite detection by PIXIE.
However, the distortions decrease with $r$, so that for $r\simeq 10^{-3}$ the spectral distortions reduce to the level similar to the standard power spectrum without running, i.e., $\mu \simeq 10^{-8}$ and $y \simeq 10^{-9}$ \citep{Chluba2012}. 
Therefore, we have an interesting cross-check between GW detection and CMB distortions. In the optimistic case, if the model of inflation realized in nature is of this type, then we expect PIXIE to detect both tensor perturbations and CMB distortions. 
\\

\section{Discussion and conclusion}
\label{sec:disc_conc}
We considered constraints on the small-scale power spectrum derived from present and future measurements of the spectral distortions in the CMB.
We introduced $k$-space window functions for $\mu$- and $y$-distortions that account for the effect of thermalization and dissipation physics and facilitate computing the effective $\mu$- and $y$-parameters directly from the primordial power spectrum at $k\gtrsim 1\,{\rm Mpc^{-1}}$ (see Eq.~\eqref{eq:mu_y_est_gen} in \S~\ref{sec:sharp}).
This defines an integral constraint that places tight limits on the shape of the small-scale power spectrum and, by extension, constrains possible inflation scenarios or any other model of the early Universe that is invoked to create the primordial seeds of the structures we see today.

We discussed different generic cases for the small-scale power spectrum, demonstrating how this integral constraint can be translated into limits on power spectrum parameters.
In particular, we derived limits from the COBE/FIRAS bounds on $\mu$ and $y$, showing that for $1\,{\rm Mpc^{-1}}\lesssim k \lesssim 10^4\,{\rm Mpc^{-1}}$ these upper limits on the amplitude of the power spectrum are roughly $10^3$ times stronger than those derived from PBHs at similar scales (see \S~\ref{sec:PBH_limits}).
Limits obtained with UCMHs supersede the COBE/FIRAS limits by more than an order of magnitude, but these limits depend on the properties of the dark matter particle.
In contrast, the constraints from CMB distortions can be obtained in a very model-independent way, relying on well-understood physics related to the thermalization and dissipation of acoustic modes.
We also showed that PIXIE will improve the bounds derived from $\mu$ and $y$ by about three orders of magnitude.  PIXIE could therefore open a new window to the early Universe, extending the lever arm from CMB-anisotropy and LSS scales all the way to $k\simeq 10^3\,{\rm Mpc^{-1}}-10^4\,{\rm Mpc^{-1}}$.

As explicit examples, we studied the constraints on inflation models with episodes of particle production (\S~\ref{sec:Neil}) and running inflaton mass (\S~\ref{sec:running_mass}).
We demonstrated that PIXIE could complement the upper limits on particle production derived at CMB and LSS scales, extending them from $k\lesssim 1\,{\rm Mpc^{-1}}$ up to $k\simeq 10^{4}\,{\rm Mpc^{-1}}$ (see Fig.~\ref{fig:constraint_particles}).
We also showed that PIXIE might have the opportunity to rule out running-mass inflation models if no spectral distortion at the level of $\mu\sim \pot{2}{-8}$ is found (see Fig.~\ref{fig:constraint_running_mass} and \S~\ref{sec:running_mass}).
 Similarly, for other models with flaring small-scale power spectrum \citep[e.g.][]{Barnaby2011} our computations indicate that strong bounds could be placed on the viable parameter space.
As argued in \S~\ref{sec:Ido}, small-field inflation models ($\Delta \phi<M_{\rm Pl}$ during inflation) with significant GW signal should simultaneously produce large CMB spectral distortions, an intriguing connection that could potentially be established by PIXIE.

The possible limits derived from future PIXIE measurements of $y$ suffer from confusion with $y$-type distortions created at low redshifts because significant differences between these two signal are not expected.
However, one effect might help in this respect: as shown by \citet{Chluba2008c} energy release before recombination causes uncompensated cycles of atomic transitions in helium and hydrogen.
This leads to extra emission and absorption features in the cosmological recombination radiation \citep{Chluba2006b, Sunyaev2009}.
At both high and very low frequencies, the associated effect is larger or comparable to the $y$-distortion itself, so that a delicate interplay between the thermalization process and the recombination radiation is expected. 
This effect might distinguish $y$-type distortions imprinted before the end of recombination from those coming from low redshifts, if precise spectral measurements are performed \citep{Chluba2008c}.
In particular, distortions at high frequencies, created by the Lyman-series of hydrogen and doubly ionized helium, as well as  the $\HeIlevel{n}{1}{P}{1}-\HeIlevel{1}{1}{S}{0}$ series of neutral helium might be very interesting in this respect.
However, to give a definite answer a more detailed computation is required, simultaneously including the effect of atomic transitions in the thermalization calculation.

The derived bounds are obtained under the assumption that the only process causing energy release at high redshifts is the dissipation of acoustic modes.
However, other forms of energy release are possible.
For example, as recently shown by \citet{Chluba2011therm} and \citet{Khatri2011}, the adiabatic cooling of ordinary matter inevitably leads to small {\it negative} $\mu$- and $y$-type distortions with amplitude $\mu \simeq -\pot{2.4}{-9}$ and $y\simeq -\text{4.3}\times 10^{-10}$.
This process is based on well understood physics, and hence the associated distortion can be predicted with high precision.
Nevertheless, many other possible sources of energy release exist
including {\it annihilating} or {\it decaying} particles \citep[e.g., see][]{Burigana1991, Hu1993, Hu1993b, McDonald2001}; {\it evaporating black holes} \citep[see][ and references therein]{Carr2010}; {\it superconducting strings} \citep{Ostriker1987, Tashiro2012}; or dissipation of {\it magnetic fields} \citep{Jedamzik2000}.
(See \citet{Chluba2011therm} for detailed computations of the associated distortions with {\sc CosmoTherm.})

All these processes come with significant uncertainties, so it is unclear at which level distortions can be expected.
Therefore, the constraints obtained here should be considered as the most {\it conservative} upper limits, since any additional energy release not caused by the dissipation of acoustic modes will only tighten the bounds on the primordial power spectrum. An interpretation of a detection of CMB spectral distortion therefore requires more careful consideration of the differences (e.g., the mixture between $\mu$ and $y$; the detailed shape of the distortion at low frequencies) in the distortions for each case,
 which in principle can be accurately computed using {\sc CosmoTherm}.
In addition, differences in the spatial distribution, although expected to be tiny for primordial distortions \citep{Chluba2012}, might help distinguish different scenarios in the future.
Correlations of the distortion with CMB anisotropies could furthermore reveal non-Gaussianity of the power spectrum \citep{Pajer2012}.

We close by mentioning that even if the parameters describing the power spectrum at CMB-anisotropy and LSS scales fully determine the small-scale power spectrum, a measurement of $\mu$ and $y$ could be interpreted as an independent confirmation of these values.
Moreover, it is important to note that the determination of $\nS$ and $\nrun$ with CMB measurements is subject to uncertainties in recombination dynamics \citep{Shaw2011}. While standard recombination physics seems to be under control \citep[e.g., see][]{Dubrovich2005, Kholupenko2007, Switzer2007II, Wong2008, Fendt2009, Jose2010, Grin2009, Chluba2010b, Yacine2010c}, possible surprises due to neglected standard or non-standard processes could still await us. Directly constraining the recombination dynamics from CMB anisotropy measurements itself is challenging \citep{Farhang2011}, so some theoretical uncertainty in the values of $\nS$ and $\nrun$ is unavoidable. 
Therefore, a detection of $\mu$-distortions at the level extrapolated from CMB and LSS scale constraints would be very reassuring, further demonstrating the great potential of this new window to the early Universe.
\\

\small

{\it Acknowledgements.} The authors are grateful to Neil Barnaby and Eric Switzer for discussions and comments on the manuscript.
Research at CITA is supported by NSERC.  ALE and IBD are also supported by the Perimeter Institute for Theoretical Physics and the Canadian Institute for Advanced Research.  Research at the Perimeter Institute is supported by the Government of Canada through Industry Canada and by the Province of Ontario through the Ministry of Research and Innovation. 
The authors also acknowledge the use of the GPC supercomputer at the SciNet HPC Consortium. SciNet is funded by: the Canada Foundation for Innovation under the auspices of Compute Canada; the Government of Ontario; Ontario Research Fund - Research Excellence; and the University of Toronto.

\bibliographystyle{apj}
\bibliography{Lit}

\begin{thebibliography}{141}
\expandafter\ifx\csname natexlab\endcsname\relax\def\natexlab#1{#1}\fi

\bibitem[{{Ach{\'u}carro} {et~al.}(2011){Ach{\'u}carro}, {Gong}, {Hardeman},
  {Palma}, \& {Patil}}]{2011JCAP...01..030A}
{Ach{\'u}carro}, A., {Gong}, J.-O., {Hardeman}, S., {Palma}, G.~A., \& {Patil},
  S.~P. 2011, \jcap, 1, 30

\bibitem[{{Adams} {et~al.}(1997){Adams}, {Ross}, \&
  {Sarkar}}]{1997NuPhB.503..405A}
{Adams}, J.~A., {Ross}, G.~G., \& {Sarkar}, S. 1997, Nuclear Physics B, 503,
  405

\bibitem[{Albrecht \& Steinhardt(1982)}]{AS82}
Albrecht, A.~J., \& Steinhardt, P.~J. 1982, \prl, 48, 1220

\bibitem[{{Ali-Ha{\"i}moud} \& {Hirata}(2011)}]{Yacine2010c}
{Ali-Ha{\"i}moud}, Y., \& {Hirata}, C.~M. 2011, \prd, 83, 043513

\bibitem[{{Atwood et al.}(2009)}]{Atwood2009}
{Atwood et al.} 2009, \apj, 697, 1071

\bibitem[{{Barnaby}(2010)}]{Barnaby2010}
{Barnaby}, N. 2010, \prd, 82, 106009

\bibitem[{{Barnaby} \& {Huang}(2009)}]{Neil2009}
{Barnaby}, N., \& {Huang}, Z. 2009, \prd, 80, 126018

\bibitem[{{Barnaby} {et~al.}(2009){Barnaby}, {Huang}, {Kofman}, \&
  {Pogosyan}}]{Neil2009I}
{Barnaby}, N., {Huang}, Z., {Kofman}, L., \& {Pogosyan}, D. 2009, \prd, 80,
  043501

\bibitem[{{Barnaby} {et~al.}(2011){Barnaby}, {Pajer}, \&
  {Peloso}}]{Barnaby2011}
{Barnaby}, N., {Pajer}, E., \& {Peloso}, M. 2011, ArXiv:1110.3327

\bibitem[{{Barrow} \& {Coles}(1991)}]{Barrow1991}
{Barrow}, J.~D., \& {Coles}, P. 1991, \mnras, 248, 52

\bibitem[{{Ben-Dayan} \& {Brustein}(2010)}]{2010JCAP...09..007B}
{Ben-Dayan}, I., \& {Brustein}, R. 2010, \jcap, 9, 7

\bibitem[{Benetti {et~al.}(2011)Benetti, Lattanzi, Calabrese, \&
  Melchiorri}]{Benetti:2011rp}
Benetti, M., Lattanzi, M., Calabrese, E., \& Melchiorri, A. 2011, Phys.Rev.,
  D84, 063509

\bibitem[{Bennett {et~al.}(2011)Bennett, Hill, Hinshaw, Larson, Smith,
  {et~al.}}]{Bennett:2010jb}
Bennett, C., Hill, R., Hinshaw, G., Larson, D., Smith, K., {et~al.} 2011,
  Astrophys.J.Suppl., 192, 17

\bibitem[{{Bennett} {et~al.}(2003){Bennett}, {Halpern}, {Hinshaw}, {Jarosik},
  {Kogut}, {Limon}, {Meyer}, {Page}, {Spergel}, {Tucker}, {Wollack}, {Wright},
  {Barnes}, {Greason}, {Hill}, {Komatsu}, {Nolta}, {Odegard}, {Peiris}, \&
  {Verde}}]{WMAP_params}
{Bennett}, C.~L., {et~al.} 2003, \apjs, 148, 1

\bibitem[{{Berezinsky} {et~al.}(2010){Berezinsky}, {Dokuchaev}, {Eroshenko},
  {Kachelrie{\ss}}, \& {Solberg}}]{BDEK10}
{Berezinsky}, V., {Dokuchaev}, V., {Eroshenko}, Y., {Kachelrie{\ss}}, M., \&
  {Solberg}, M.~A. 2010, \prd, 81, 103529

\bibitem[{{Bird} {et~al.}(2011){Bird}, {Peiris}, {Viel}, \& {Verde}}]{BPVV11}
{Bird}, S., {Peiris}, H.~V., {Viel}, M., \& {Verde}, L. 2011, \mnras, 413, 1717

\bibitem[{{Bringmann} {et~al.}(2011){Bringmann}, {Scott}, \& {Akrami}}]{BSA11}
{Bringmann}, T., {Scott}, P., \& {Akrami}, Y. 2011, ArXiv:1110.2484v1

\bibitem[{{Brown} {et~al.}(2009){Brown}, {Ade}, {Bock}, {Bowden}, {Cahill},
  {Castro}, {Church}, {Culverhouse}, {Friedman}, {Ganga}, {Gear}, {Gupta},
  {Hinderks}, {Kovac}, {Lange}, {Leitch}, {Melhuish}, {Memari}, {Murphy},
  {Orlando}, {O'Sullivan}, {Piccirillo}, {Pryke}, {Rajguru}, {Rusholme},
  {Schwarz}, {Taylor}, {Thompson}, {Turner}, {Wu}, {Zemcov}, \& {The QUa D
  collaboration}}]{Quad09}
{Brown}, M.~L., {et~al.} 2009, \apj, 705, 978

\bibitem[{{Bugaev} \& {Klimai}(2011{\natexlab{a}})}]{2011JCAP...11..028B}
{Bugaev}, E., \& {Klimai}, P. 2011{\natexlab{a}}, \jcap, 11, 28

\bibitem[{{Bugaev} \& {Klimai}(2011{\natexlab{b}})}]{2011arXiv1112.5601B}
---. 2011{\natexlab{b}}, ArXiv:1112.5601

\bibitem[{{Burigana} {et~al.}(1991){Burigana}, {Danese}, \& {de
  Zotti}}]{Burigana1991}
{Burigana}, C., {Danese}, L., \& {de Zotti}, G. 1991, \aap, 246, 49

\bibitem[{{Carr}(1975)}]{Carr75}
{Carr}, B.~J. 1975, \apj, 201, 1

\bibitem[{{Carr} {et~al.}(2010){Carr}, {Kohri}, {Sendouda}, \&
  {Yokoyama}}]{Carr2010}
{Carr}, B.~J., {Kohri}, K., {Sendouda}, Y., \& {Yokoyama}, J. 2010, \prd, 81,
  104019

\bibitem[{{Carr} \& {Lidsey}(1993)}]{1993PhRvD..48..543C}
{Carr}, B.~J., \& {Lidsey}, J.~E. 1993, \prd, 48, 543

\bibitem[{{Cen} \& {Ostriker}(1999)}]{Cen1999}
{Cen}, R., \& {Ostriker}, J.~P. 1999, \apj, 514, 1

\bibitem[{{Cespedes} {et~al.}(2012){Cespedes}, {Atal}, \&
  {Palma}}]{2012arXiv1201.4848C}
{Cespedes}, S., {Atal}, V., \& {Palma}, G.~A. 2012, ArXiv:1201.4848

\bibitem[{{Chluba} {et~al.}(2012){Chluba}, {Khatri}, \& {Sunyaev}}]{Chluba2012}
{Chluba}, J., {Khatri}, R., \& {Sunyaev}, R.~A. 2012, ArXiv:1202.0057

\bibitem[{{Chluba} \& {Sunyaev}(2004)}]{Chluba2004}
{Chluba}, J., \& {Sunyaev}, R.~A. 2004, \aap, 424, 389

\bibitem[{{Chluba} \& {Sunyaev}(2006)}]{Chluba2006b}
---. 2006, \aap, 458, L29

\bibitem[{{Chluba} \& {Sunyaev}(2009)}]{Chluba2008c}
---. 2009, \aap, 501, 29

\bibitem[{{Chluba} \& {Sunyaev}(2012)}]{Chluba2011therm}
---. 2012, \mnras, 419, 1294

\bibitem[{{Chluba} \& {Thomas}(2011)}]{Chluba2010b}
{Chluba}, J., \& {Thomas}, R.~M. 2011, \mnras, 412, 748

\bibitem[{{Chung} {et~al.}(2000){Chung}, {Kolb}, {Riotto}, \&
  {Tkachev}}]{2000PhRvD..62d3508C}
{Chung}, D.~J.~H., {Kolb}, E.~W., {Riotto}, A., \& {Tkachev}, I.~I. 2000, \prd,
  62, 043508

\bibitem[{{Copeland} {et~al.}(1998){Copeland}, {Liddle}, {Lidsey}, \&
  {Wands}}]{1998PhRvD..58f3508C}
{Copeland}, E.~J., {Liddle}, A.~R., {Lidsey}, J.~E., \& {Wands}, D. 1998, \prd,
  58, 063508

\bibitem[{{Covi} \& {Lyth}(1999)}]{1999PhRvD..59f3515C}
{Covi}, L., \& {Lyth}, D.~H. 1999, \prd, 59, 063515

\bibitem[{{Covi} {et~al.}(2004){Covi}, {Lyth}, {Melchiorri}, \&
  {Odman}}]{Covi2004}
{Covi}, L., {Lyth}, D.~H., {Melchiorri}, A., \& {Odman}, C.~J. 2004, \prd, 70,
  123521

\bibitem[{{Covi} {et~al.}(1999){Covi}, {Lyth}, \&
  {Roszkowski}}]{1999PhRvD..60b3509C}
{Covi}, L., {Lyth}, D.~H., \& {Roszkowski}, L. 1999, \prd, 60, 023509

\bibitem[{{Daly}(1991)}]{Daly1991}
{Daly}, R.~A. 1991, \apj, 371, 14

\bibitem[{{Dent} {et~al.}(2012){Dent}, {Easson}, \& {Tashiro}}]{Dent2012}
{Dent}, J.~B., {Easson}, D.~A., \& {Tashiro}, H. 2012, ArXiv:1202.6066

\bibitem[{{Dubrovich} \& {Grachev}(2005)}]{Dubrovich2005}
{Dubrovich}, V.~K., \& {Grachev}, S.~I. 2005, Astronomy Letters, 31, 359

\bibitem[{{Dunkley} {et~al.}(2011){Dunkley}, {Hlozek}, {Sievers}, {Acquaviva},
  {Ade}, {Aguirre}, {Amiri}, {Appel}, {Barrientos}, {Battistelli}, \&
  {Bond}}]{Dunkley2010}
{Dunkley}, J., {et~al.} 2011, \apj, 739, 52

\bibitem[{{Dvorkin} \& {Hu}(2010)}]{Dvorkin2010}
{Dvorkin}, C., \& {Hu}, W. 2010, \prd, 82, 043513

\bibitem[{{Dvorkin} \& {Hu}(2011)}]{Dvorkin2011}
---. 2011, \prd, 84, 063515

\bibitem[{{Farhang} {et~al.}(2011){Farhang}, {Bond}, \& {Chluba}}]{Farhang2011}
{Farhang}, M., {Bond}, J.~R., \& {Chluba}, J. 2011, ArXiv:1110.4608

\bibitem[{{Fendt} {et~al.}(2009){Fendt}, {Chluba}, {Rubi{\~n}o-Mart{\'{\i}}n},
  \& {Wandelt}}]{Fendt2009}
{Fendt}, W.~A., {Chluba}, J., {Rubi{\~n}o-Mart{\'{\i}}n}, J.~A., \& {Wandelt},
  B.~D. 2009, \apjs, 181, 627

\bibitem[{{Fixsen} {et~al.}(1996){Fixsen}, {Cheng}, {Gales}, {Mather},
  {Shafer}, \& {Wright}}]{Fixsen1996}
{Fixsen}, D.~J., {Cheng}, E.~S., {Gales}, J.~M., {Mather}, J.~C., {Shafer},
  R.~A., \& {Wright}, E.~L. 1996, \apj, 473, 576

\bibitem[{{Gervasi} {et~al.}(2008){Gervasi}, {Zannoni}, {Tartari}, {Boella}, \&
  {Sironi}}]{tris2}
{Gervasi}, M., {Zannoni}, M., {Tartari}, A., {Boella}, G., \& {Sironi}, G.
  2008, \apj, 688, 24

\bibitem[{{Gong} \& {Sasaki}(2011)}]{2011JCAP...03..028G}
{Gong}, J.-O., \& {Sasaki}, M. 2011, \jcap, 3, 28

\bibitem[{{Grin} \& {Hirata}(2010)}]{Grin2009}
{Grin}, D., \& {Hirata}, C.~M. 2010, \prd, 81, 083005

\bibitem[{Guth(1981)}]{Guth80}
Guth, A.~H. 1981, \prd, 23, 347

\bibitem[{Hamann {et~al.}(2010)Hamann, Shafieloo, \& Souradeep}]{Hamann:2009bz}
Hamann, J., Shafieloo, A., \& Souradeep, T. 2010, JCAP, 1004, 010

\bibitem[{{Hlozek} {et~al.}(2011){Hlozek}, {Dunkley}, {Addison}, {Appel},
  {Bond}, {Sofia Carvalho}, {Das}, {Devlin}, {D{\"u}nner}, {Essinger-Hileman},
  {Fowler}, {Gallardo}, {Hajian}, {Halpern}, {Hasselfield}, {Hilton}, {Hincks},
  {Hughes}, {Irwin}, {Klein}, {Kosowsky}, {Marriage}, {Marsden}, {Menanteau},
  {Moodley}, {Niemack}, {Nolta}, {Page}, {Parker}, {Partridge}, {Rojas},
  {Sehgal}, {Sherwin}, {Sievers}, {Spergel}, {Staggs}, {Swetz}, {Switzer},
  {Thornton}, \& {Wollack}}]{ACT11}
{Hlozek}, R., {et~al.} 2011, ArXiv:1105.4887

\bibitem[{{Hotchkiss} {et~al.}(2012){Hotchkiss}, {Mazumdar}, \&
  {Nadathur}}]{Hotchkiss2012}
{Hotchkiss}, S., {Mazumdar}, A., \& {Nadathur}, S. 2012, \jcap, 2, 8

\bibitem[{{Hu} {et~al.}(1994){Hu}, {Scott}, \& {Silk}}]{Hu1994}
{Hu}, W., {Scott}, D., \& {Silk}, J. 1994, \apjl, 430, L5

\bibitem[{{Hu} \& {Silk}(1993{\natexlab{a}})}]{Hu1993}
{Hu}, W., \& {Silk}, J. 1993{\natexlab{a}}, \prd, 48, 485

\bibitem[{{Hu} \& {Silk}(1993{\natexlab{b}})}]{Hu1993b}
---. 1993{\natexlab{b}}, Physical Review Letters, 70, 2661

\bibitem[{{Hu} \& {Sugiyama}(1994)}]{Hu1994isocurv}
{Hu}, W., \& {Sugiyama}, N. 1994, \apj, 436, 456

\bibitem[{Hunt \& Sarkar(2007)}]{Hunt:2007dn}
Hunt, P., \& Sarkar, S. 2007, Phys.Rev., D76, 123504

\bibitem[{{Illarionov} \& {Sunyaev}(1974)}]{Illarionov1974}
{Illarionov}, A.~F., \& {Sunyaev}, R.~A. 1974, Astronomicheskii Zhurnal, 51,
  1162

\bibitem[{{Ivanov} {et~al.}(1994){Ivanov}, {Naselsky}, \&
  {Novikov}}]{1994PhRvD..50.7173I}
{Ivanov}, P., {Naselsky}, P., \& {Novikov}, I. 1994, \prd, 50, 7173

\bibitem[{{Jedamzik} {et~al.}(2000){Jedamzik}, {Katalini{\'c}}, \&
  {Olinto}}]{Jedamzik2000}
{Jedamzik}, K., {Katalini{\'c}}, V., \& {Olinto}, A.~V. 2000, Physical Review
  Letters, 85, 700

\bibitem[{{Josan} \& {Green}(2010{\natexlab{a}})}]{2010PhRvD..82d7303J}
{Josan}, A.~S., \& {Green}, A.~M. 2010{\natexlab{a}}, \prd, 82, 047303

\bibitem[{{Josan} \& {Green}(2010{\natexlab{b}})}]{JG10}
---. 2010{\natexlab{b}}, \prd, 82, 083527

\bibitem[{{Josan} {et~al.}(2009){Josan}, {Green}, \& {Malik}}]{JGM09}
{Josan}, A.~S., {Green}, A.~M., \& {Malik}, K.~A. 2009, \prd, 79, 103520

\bibitem[{{Joy} {et~al.}(2008){Joy}, {Sahni}, \&
  {Starobinsky}}]{2008PhRvD..77b3514J}
{Joy}, M., {Sahni}, V., \& {Starobinsky}, A.~A. 2008, \prd, 77, 023514

\bibitem[{{Kaiser}(1983)}]{Kaiser1983}
{Kaiser}, N. 1983, \mnras, 202, 1169

\bibitem[{{Keisler} {et~al.}(2011){Keisler}, {Reichardt}, {Aird}, {Benson},
  {Bleem}, {Carlstrom}, {Chang}, {Cho}, {Crawford}, {Crites}, {de Haan},
  {Dobbs}, {Dudley}, {George}, \& {Halverson}}]{Keisler2011}
{Keisler}, R., {et~al.} 2011, \apj, 743, 28

\bibitem[{{Khatri} {et~al.}(2011){Khatri}, {Sunyaev}, \& {Chluba}}]{Khatri2011}
{Khatri}, R., {Sunyaev}, R.~A., \& {Chluba}, J. 2011, arXiv:1110.0475

\bibitem[{{Kholupenko} {et~al.}(2007){Kholupenko}, {Ivanchik}, \&
  {Varshalovich}}]{Kholupenko2007}
{Kholupenko}, E.~E., {Ivanchik}, A.~V., \& {Varshalovich}, D.~A. 2007, \mnras,
  378, L39

\bibitem[{Kinney {et~al.}(2008)Kinney, Kolb, Melchiorri, \&
  Riotto}]{Kinney:2008wy}
Kinney, W.~H., Kolb, E.~W., Melchiorri, A., \& Riotto, A. 2008, Phys.Rev., D78,
  087302

\bibitem[{{Kobayashi} \& {Takahashi}(2011)}]{Takeshi2011}
{Kobayashi}, T., \& {Takahashi}, F. 2011, \jcap, 1, 26

\bibitem[{{Kogut} {et~al.}(2011){Kogut}, {Fixsen}, {Chuss}, {Dotson}, {Dwek},
  {Halpern}, {Hinshaw}, {Meyer}, {Moseley}, {Seiffert}, {Spergel}, \&
  {Wollack}}]{Kogut2011PIXIE}
{Kogut}, A., {et~al.} 2011, \jcap, 7, 25

\bibitem[{{Kohri} {et~al.}(2008){Kohri}, {Lyth}, \&
  {Melchiorri}}]{2008JCAP...04..038K}
{Kohri}, K., {Lyth}, D.~H., \& {Melchiorri}, A. 2008, \jcap, 4, 38

\bibitem[{{Komatsu} {et~al.}(2011){Komatsu}, {Smith}, {Dunkley}, {Bennett},
  {Gold}, {Hinshaw}, {Jarosik}, {Larson}, {Nolta}, {Page}, {Spergel},
  {Halpern}, {Hill}, {Kogut}, {Limon}, {Meyer}, {Odegard}, \&
  {Tucker}}]{Komatsu2010}
{Komatsu}, E., {et~al.} 2011, \apjs, 192, 18

\bibitem[{{Kosowsky} \& {Turner}(1995)}]{Kosowsky1995}
{Kosowsky}, A., \& {Turner}, M.~S. 1995, \prd, 52, 1739

\bibitem[{{Lacki} \& {Beacom}(2010)}]{LB10}
{Lacki}, B.~C., \& {Beacom}, J.~F. 2010, \apjl, 720, L67

\bibitem[{{Larson} {et~al.}(2011){Larson}, {Dunkley}, {Hinshaw}, {Komatsu},
  {Nolta}, {Bennett}, {Gold}, {Halpern}, {Hill}, {Jarosik}, {Kogut}, {Limon},
  {Meyer}, {Odegard}, \& {Page}}]{Larson2011}
{Larson}, D., {et~al.} 2011, \apjs, 192, 16

\bibitem[{{Leach} {et~al.}(2000){Leach}, {Grivell}, \&
  {Liddle}}]{2000PhRvD..62d3516L}
{Leach}, S.~M., {Grivell}, I.~J., \& {Liddle}, A.~R. 2000, \prd, 62, 043516

\bibitem[{{Li} {et~al.}(2012){Li}, {Erickcek}, \& {Law}}]{LEW12}
{Li}, F., {Erickcek}, A.~L., \& {Law}, N.~M. 2012, ArXiv:1202.1284

\bibitem[{{Lidsey} {et~al.}(1997){Lidsey}, {Liddle}, {Kolb}, {Copeland},
  {Barreiro}, \& {Abney}}]{LLK97}
{Lidsey}, J.~E., {Liddle}, A.~R., {Kolb}, E.~W., {Copeland}, E.~J., {Barreiro},
  T., \& {Abney}, M. 1997, Reviews of Modern Physics, 69, 373

\bibitem[{Linde(1982)}]{Linde82}
Linde, A.~D. 1982, \plb, 108, 389

\bibitem[{{Lindegren} {et~al.}(2012){Lindegren}, {Lammers}, {Hobbs},
  {O'Mullane}, {Bastian}, \& {Hern{\'a}ndez}}]{Lindegren2011}
{Lindegren}, L., {Lammers}, U., {Hobbs}, D., {O'Mullane}, W., {Bastian}, U., \&
  {Hern{\'a}ndez}, J. 2012, A\&A in press

\bibitem[{{Lyth}(2011{\natexlab{a}})}]{2011JCAP...07..035L}
{Lyth}, D.~H. 2011{\natexlab{a}}, \jcap, 7, 35

\bibitem[{{Lyth}(2011{\natexlab{b}})}]{2011arXiv1107.1681L}
---. 2011{\natexlab{b}}, ArXiv:1107.1681

\bibitem[{{Lyth} \& {Riotto}(1999)}]{Lyth1999}
{Lyth}, D.~H.~D.~H., \& {Riotto}, A.~A. 1999, \physrep, 314, 1

\bibitem[{{Markevitch} {et~al.}(1991){Markevitch}, {Blumenthal}, {Forman},
  {Jones}, \& {Sunyaev}}]{Markevitch1991}
{Markevitch}, M., {Blumenthal}, G.~R., {Forman}, W., {Jones}, C., \& {Sunyaev},
  R.~A. 1991, \apjl, 378, L33

\bibitem[{{Martin} \& {Brandenberger}(2001)}]{2001PhRvD..63l3501M}
{Martin}, J., \& {Brandenberger}, R.~H. 2001, \prd, 63, 123501

\bibitem[{{Martin} {et~al.}(2000){Martin}, {Riazuelo}, \&
  {Sakellariadou}}]{2000PhRvD..61h3518M}
{Martin}, J., {Riazuelo}, A., \& {Sakellariadou}, M. 2000, \prd, 61, 083518

\bibitem[{{Mather} {et~al.}(1994){Mather}, {Cheng}, {Cottingham}, {Eplee},
  {Fixsen}, {Hewagama}, {Isaacman}, {Jensen}, {Meyer}, {Noerdlinger}, {Read},
  \& {Rosen}}]{Mather1994}
{Mather}, J.~C., {et~al.} 1994, \apj, 420, 439

\bibitem[{{McDonald} {et~al.}(2001){McDonald}, {Scherrer}, \&
  {Walker}}]{McDonald2001}
{McDonald}, P., {Scherrer}, R.~J., \& {Walker}, T.~P. 2001, \prd, 63, 023001

\bibitem[{{McDonald} {et~al.}(2006){McDonald}, {Seljak}, {Burles}, {Schlegel},
  {Weinberg}, {Cen}, {Shih}, {Schaye}, {Schneider}, {Bahcall}, {Briggs},
  {Brinkmann}, {Brunner}, {Fukugita}, {Gunn}, {Ivezi{\'c}}, {Kent}, {Lupton},
  \& {Vanden Berk}}]{MSB06}
{McDonald}, P., {et~al.} 2006, \apjs, 163, 80

\bibitem[{{McQuinn} {et~al.}(2005){McQuinn}, {Furlanetto}, {Hernquist}, {Zahn},
  \& {Zaldarriaga}}]{McQuinn2005}
{McQuinn}, M., {Furlanetto}, S.~R., {Hernquist}, L., {Zahn}, O., \&
  {Zaldarriaga}, M. 2005, \apj, 630, 643

\bibitem[{{Miniati} {et~al.}(2000){Miniati}, {Ryu}, {Kang}, {Jones}, {Cen}, \&
  {Ostriker}}]{Miniati2000}
{Miniati}, F., {Ryu}, D., {Kang}, H., {Jones}, T.~W., {Cen}, R., \& {Ostriker},
  J.~P. 2000, \apj, 542, 608

\bibitem[{Mortonson {et~al.}(2009)Mortonson, Dvorkin, Peiris, \&
  Hu}]{Mortonson:2009qv}
Mortonson, M.~J., Dvorkin, C., Peiris, H.~V., \& Hu, W. 2009, Phys.Rev., D79,
  103519

\bibitem[{{Nicholson} \& {Contaldi}(2009)}]{NC09}
{Nicholson}, G., \& {Contaldi}, C.~R. 2009, \jcap, 7, 11

\bibitem[{{Niemeyer} \& {Jedamzik}(1999)}]{NJ99}
{Niemeyer}, J.~C., \& {Jedamzik}, K. 1999, \prd, 59, 124013

\bibitem[{{Oh} {et~al.}(2003){Oh}, {Cooray}, \& {Kamionkowski}}]{Oh2003}
{Oh}, S.~P., {Cooray}, A., \& {Kamionkowski}, M. 2003, \mnras, 342, L20

\bibitem[{{Ostriker} \& {Thompson}(1987)}]{Ostriker1987}
{Ostriker}, J.~P., \& {Thompson}, C. 1987, \apjl, 323, L97

\bibitem[{{Pajer} \& {Zaldarriaga}(2012)}]{Pajer2012}
{Pajer}, E., \& {Zaldarriaga}, M. 2012, ArXiv:1201.5375

\bibitem[{{Pearson} {et~al.}(2003){Pearson}, {Mason}, {Readhead}, {Shepherd},
  {Sievers}, {Udomprasert}, {Cartwright}, {Farmer}, {Padin}, {Myers}, {Bond},
  {Contaldi}, {Pen}, {Prunet}, {Pogosyan}, {Carlstrom}, {Kovac}, {Leitch},
  {Pryke}, {Halverson}, {Holzapfel}, {Altamirano}, {Bronfman}, {Casassus},
  {May}, \& {Joy}}]{CBI03}
{Pearson}, T.~J., {et~al.} 2003, \apj, 591, 556

\bibitem[{{Peiris} \& {Easther}(2008)}]{2008JCAP...07..024P}
{Peiris}, H.~V., \& {Easther}, R. 2008, \jcap, 7, 24

\bibitem[{Peiris \& Verde(2010)}]{Peiris:2009wp}
Peiris, H.~V., \& Verde, L. 2010, Phys.Rev., D81, 021302

\bibitem[{Polarski \& Starobinsky(1992)}]{Polarski:1992dq}
Polarski, D., \& Starobinsky, A.~A. 1992, Nucl. Phys., B385, 623

\bibitem[{{Randall} {et~al.}(1996){Randall}, {Solja{\v c}i{\'C}}, \&
  {Guth}}]{1996NuPhB.472..377R}
{Randall}, L., {Solja{\v c}i{\'C}}, M., \& {Guth}, A.~H. 1996, Nuclear Physics
  B, 472, 377

\bibitem[{{Reichardt} {et~al.}(2009){Reichardt}, {Ade}, {Bock}, {Bond},
  {Brevik}, {Contaldi}, {Daub}, {Dempsey}, {Goldstein}, {Holzapfel}, {Kuo},
  {Lange}, {Lueker}, {Newcomb}, {Peterson}, {Ruhl}, {Runyan}, \&
  {Staniszewski}}]{ACBAR09}
{Reichardt}, C.~L., {et~al.} 2009, \apj, 694, 1200

\bibitem[{{Reid} {et~al.}(2010){Reid}, {Percival}, {Eisenstein}, {Verde},
  {Spergel}, {Skibba}, {Bahcall}, {Budavari}, {Frieman}, {Fukugita}, {Gott},
  {Gunn}, {Ivezi{\'c}}, {Knapp}, {Kron}, {Lupton}, {McKay}, {Meiksin},
  {Nichol}, {Pope}, {Schlegel}, {Schneider}, {Stoughton}, {Strauss}, {Szalay},
  {Tegmark}, {Vogeley}, {Weinberg}, {York}, \& {Zehavi}}]{SDSSDR7}
{Reid}, B.~A., {et~al.} 2010, \mnras, 404, 60

\bibitem[{{Ricotti} \& {Gould}(2009)}]{RG09}
{Ricotti}, M., \& {Gould}, A. 2009, \apj, 707, 979

\bibitem[{{Rubi{\~n}o-Mart{\'{\i}}n} {et~al.}(2010){Rubi{\~n}o-Mart{\'{\i}}n},
  {Chluba}, {Fendt}, \& {Wandelt}}]{Jose2010}
{Rubi{\~n}o-Mart{\'{\i}}n}, J.~A., {Chluba}, J., {Fendt}, W.~A., \& {Wandelt},
  B.~D. 2010, \mnras, 403, 439

\bibitem[{{Salopek} {et~al.}(1989){Salopek}, {Bond}, \&
  {Bardeen}}]{1989PhRvD..40.1753S}
{Salopek}, D.~S., {Bond}, J.~R., \& {Bardeen}, J.~M. 1989, \prd, 40, 1753

\bibitem[{{Scott} \& {Sivertsson}(2009)}]{SS09}
{Scott}, P., \& {Sivertsson}, S. 2009, Physical Review Letters, 103, 211301

\bibitem[{{Sehgal} {et~al.}(2011){Sehgal}, {Trac}, {Acquaviva}, {Ade},
  {Aguirre}, {Amiri}, {Appel}, {Barrientos}, {Battistelli}, {Bond}, {Brown},
  {Burger}, {Chervenak}, {Das}, {Devlin}, {Dicker}, {Bertrand Doriese},
  {Dunkley}, {D{\"u}nner}, {Essinger-Hileman}, {Fisher}, {Fowler}, {Hajian},
  {Halpern}, {Hasselfield}, {Hern{\'a}ndez-Monteagudo}, {Hilton}, {Hilton},
  {Hincks}, {Hlozek}, {Holtz}, {Huffenberger}, {Hughes}, {Hughes}, {Infante},
  {Irwin}, {Jones}, {Baptiste Juin}, {Klein}, {Kosowsky}, {Lau}, {Limon},
  {Lin}, {Lupton}, {Marriage}, {Marsden}, {Martocci}, {Mauskopf}, {Menanteau},
  {Moodley}, {Moseley}, {Netterfield}, {Niemack}, {Nolta}, {Page}, {Parker},
  {Partridge}, {Reid}, {Sherwin}, {Sievers}, {Spergel}, {Staggs}, {Swetz},
  {Switzer}, {Thornton}, {Tucker}, {Warne}, {Wollack}, \& {Zhao}}]{STA11}
{Sehgal}, N., {et~al.} 2011, \apj, 732, 44

\bibitem[{{Seiffert} {et~al.}(2011){Seiffert}, {Fixsen}, {Kogut}, {Levin},
  {Limon}, {Lubin}, {Mirel}, {Singal}, {Villela}, {Wollack}, \&
  {Wuensche}}]{arcade2}
{Seiffert}, M., {et~al.} 2011, \apj, 734, 6

\bibitem[{{Seljak} {et~al.}(2005){Seljak}, {Makarov}, {McDonald}, {Anderson},
  {Bahcall}, {Brinkmann}, {Burles}, {Cen}, {Doi}, {Gunn}, {Ivezi{\'c}}, {Kent},
  {Loveday}, {Lupton}, {Munn}, {Nichol}, {Ostriker}, {Schlegel}, {Schneider},
  {Tegmark}, {Berk}, {Weinberg}, \& {York}}]{Seljak2005}
{Seljak}, U., {et~al.} 2005, \prd, 71, 103515

\bibitem[{{Shafi} \& {Wickman}(2011)}]{Shafi2011}
{Shafi}, Q., \& {Wickman}, J.~R. 2011, Physics Letters B, 696, 438

\bibitem[{{Shaw} \& {Chluba}(2011)}]{Shaw2011}
{Shaw}, J.~R., \& {Chluba}, J. 2011, \mnras, 415, 1343

\bibitem[{{Silk}(1968)}]{Silk1968}
{Silk}, J. 1968, \apj, 151, 459

\bibitem[{{Silk} \& {Turner}(1987)}]{1987PhRvD..35..419S}
{Silk}, J., \& {Turner}, M.~S. 1987, \prd, 35, 419

\bibitem[{{Starobinskij}(1992)}]{1992JETPL..55..489S}
{Starobinskij}, A.~A. 1992, Soviet Journal of Experimental and Theoretical
  Physics Letters, 55, 489

\bibitem[{{Starobinsky}(1998)}]{1998GrCo....4S..88S}
{Starobinsky}, A.~A. 1998, Gravitation and Cosmology, 4, 88

\bibitem[{{Stewart}(1997{\natexlab{a}})}]{Stewart1997}
{Stewart}, E.~D. 1997{\natexlab{a}}, Physics Letters B, 391, 34

\bibitem[{{Stewart}(1997{\natexlab{b}})}]{Stewart1997b}
---. 1997{\natexlab{b}}, \prd, 56, 2019

\bibitem[{{Sunyaev} \& {Chluba}(2009)}]{Sunyaev2009}
{Sunyaev}, R.~A., \& {Chluba}, J. 2009, Astronomische Nachrichten, 330, 657

\bibitem[{{Sunyaev} \& {Zeldovich}(1970{\natexlab{a}})}]{Sunyaev1970diss}
{Sunyaev}, R.~A., \& {Zeldovich}, Y.~B. 1970{\natexlab{a}}, \apss, 9, 368

\bibitem[{{Sunyaev} \& {Zeldovich}(1970{\natexlab{b}})}]{Sunyaev1970mu}
---. 1970{\natexlab{b}}, \apss, 7, 20

\bibitem[{{Sunyaev} \& {Zeldovich}(1972)}]{Sunyaev1972b}
---. 1972, \aap, 20, 189

\bibitem[{{Switzer} \& {Hirata}(2008)}]{Switzer2007II}
{Switzer}, E.~R., \& {Hirata}, C.~M. 2008, \prd, 77, 083008

\bibitem[{{Tashiro} {et~al.}(2012){Tashiro}, {Sabancilar}, \&
  {Vachaspati}}]{Tashiro2012}
{Tashiro}, H., {Sabancilar}, E., \& {Vachaspati}, T. 2012, ArXiv:1202.2474

\bibitem[{{Tegmark} \& {Zaldarriaga}(2002)}]{TZ02}
{Tegmark}, M., \& {Zaldarriaga}, M. 2002, \prd, 66, 103508

\bibitem[{{Tegmark} {et~al.}(2004){Tegmark}, {Strauss}, {Blanton}, {Abazajian},
  {Dodelson}, {Sandvik}, {Wang}, {Weinberg}, {Zehavi}, {Bahcall}, {Hoyle},
  {Schlegel}, {Scoccimarro}, {Vogeley}, {Berlind}, {Budavari}, {Connolly},
  {Eisenstein}, {Finkbeiner}, {Frieman}, {Gunn}, {Hui}, {Jain}, {Johnston},
  {Kent}, {Lin}, {Nakajima}, {Nichol}, {Ostriker}, {Pope}, {Scranton},
  {Seljak}, {Sheth}, {Stebbins}, {Szalay}, {Szapudi}, {Xu}, {Annis},
  {Brinkmann}, {Burles}, {Castander}, {Csabai}, {Loveday}, {Doi}, {Fukugita},
  {Gillespie}, {Hennessy}, {Hogg}, {Ivezi{\'c}}, {Knapp}, {Lamb}, {Lee},
  {Lupton}, {McKay}, {Kunszt}, {Munn}, {O'Connell}, {Peoples}, {Pier},
  {Richmond}, {Rockosi}, {Schneider}, {Stoughton}, {Tucker}, {vanden Berk},
  {Yanny}, \& {York}}]{Tegmark2004}
{Tegmark}, M., {et~al.} 2004, \prd, 69, 103501

\bibitem[{{Tinker} {et~al.}(2012){Tinker}, {Sheldon}, {Wechsler}, {Becker},
  {Rozo}, {Zu}, {Weinberg}, {Zehavi}, {Blanton}, {Busha}, \& {Koester}}]{TSW11}
{Tinker}, J.~L., {et~al.} 2012, \apj, 745, 16

\bibitem[{{Vikhlinin} {et~al.}(2009){Vikhlinin}, {Kravtsov}, {Burenin},
  {Ebeling}, {Forman}, {Hornstrup}, {Jones}, {Murray}, {Nagai}, {Quintana}, \&
  {Voevodkin}}]{VKB09}
{Vikhlinin}, A., {et~al.} 2009, \apj, 692, 1060

\bibitem[{{Weinberg}(2008)}]{WeinbergBook}
{Weinberg}, S. 2008, {Cosmology} (Oxford University Press)

\bibitem[{{Wong} {et~al.}(2008){Wong}, {Moss}, \& {Scott}}]{Wong2008}
{Wong}, W.~Y., {Moss}, A., \& {Scott}, D. 2008, \mnras, 386, 1023

\bibitem[{Yang {et~al.}(2011)Yang, Chen, Lu, \& Zong}]{YCL11}
Yang, Y., Chen, X., Lu, T., \& Zong, H. 2011, Eur. Phys. J. Plus, 126, 123

\bibitem[{{Yang} {et~al.}(2011{\natexlab{a}}){Yang}, {Feng}, {Huang}, {Chen},
  {Lu}, \& {Zong}}]{YFH11}
{Yang}, Y., {Feng}, L., {Huang}, X., {Chen}, X., {Lu}, T., \& {Zong}, H.
  2011{\natexlab{a}}, \jcap, 12, 20

\bibitem[{{Yang} {et~al.}(2011{\natexlab{b}}){Yang}, {Huang}, {Chen}, \&
  {Zong}}]{YHCZ11}
{Yang}, Y., {Huang}, X., {Chen}, X., \& {Zong}, H. 2011{\natexlab{b}}, \prd,
  84, 043506

\bibitem[{{Zaldarriaga} \& {Harari}(1995)}]{Zaldarriaga1995}
{Zaldarriaga}, M., \& {Harari}, D.~D. 1995, \prd, 52, 3276

\bibitem[{{Zannoni} {et~al.}(2008){Zannoni}, {Tartari}, {Gervasi}, {Boella},
  {Sironi}, {De Lucia}, {Passerini}, \& {Cavaliere}}]{tris1}
{Zannoni}, M., {Tartari}, A., {Gervasi}, M., {Boella}, G., {Sironi}, G., {De
  Lucia}, A., {Passerini}, A., \& {Cavaliere}, F. 2008, \apj, 688, 12

\bibitem[{{Zel'Dovich} {et~al.}(1972){Zel'Dovich}, {Illarionov}, \&
  {Syunyaev}}]{Zeldovich1972}
{Zel'Dovich}, Y.~B., {Illarionov}, A.~F., \& {Syunyaev}, R.~A. 1972, Soviet
  Journal of Experimental and Theoretical Physics, 35, 643

\bibitem[{{Zeldovich} \& {Sunyaev}(1969)}]{Zeldovich1969}
{Zeldovich}, Y.~B., \& {Sunyaev}, R.~A. 1969, \apss, 4, 301

\bibitem[{{Zhang}(2011)}]{Zhang11}
{Zhang}, D. 2011, \mnras, 418, 1850

\end{thebibliography}

\end{document}